\documentclass[12pt]{article}
\usepackage{amssymb}
\usepackage{euscript}

\let\ssection=\section
\renewcommand{\section}{\setcounter{equation}{0}\ssection}

\newcommand{\be}{\begin{equation}}
\newcommand{\ee}{\end{equation}}

\newcommand{\raw}{\rightarrow}

\newcommand\mathC{\mkern1mu\raise2.2pt\hbox{$\scriptscriptstyle|$}
		{\mkern-7mu\rm C}}				
		
\def\Dslash{\setbox0=\hbox{$D$}D\hskip-\wd0\hbox to\wd0{\hss\sl/\/\hss}}

\begin{document}
\begin{center}
{\large\bf Emergence, Reduction and Supervenience: a Varied Landscape}
\end{center}

\vspace{0.8 truecm}
\begin{center}
            J.~Butterfield
\\[10pt]
Trinity College, Cambridge University, Cambridge CB2 1TQ; email: jb56@cam.ac.uk
\end{center}


\vspace{0.6in}

\begin{abstract}
This is one of two papers about emergence, reduction and supervenience. It expounds these 
notions and analyses the general relations between them. The companion paper analyses the 
situation in physics, especially limiting relations between physical theories.

I shall take emergence as behaviour that is novel and robust relative to some comparison 
class. I shall take reduction as deduction using appropriate auxiliary definitions. And I 
shall take supervenience as a weakening of reduction, viz. to allow infinitely long 
definitions.

The overall claim of this paper will be that emergence is logically independent both of 
reduction and of supervenience. In particular, one can have emergence with reduction, as well 
as without it; and emergence without supervenience, as well as with it.

Of the subsidiary claims, the four main ones (each shared with some other authors) are:\\
\indent (i): I defend the traditional Nagelian conception of reduction (Section \ref{No1});  
\\
\indent (ii): I deny that the multiple realizability argument causes trouble for reductions, 
or ``reductionism'' (Section \ref{superdef}); \\
\indent (iii): I stress the collapse of supervenience into deduction via Beth's theorem 
(Section \ref{Beth});\\
\indent (iv): I adapt some examples already in the literature to show supervenience  without 
emergence and {\em vice versa} (Section \ref{SEindpdt}).

\end{abstract}
\newpage
\tableofcontents

\section{Introduction}\label{Intr}
This is one of a pair of papers that together rebut two philosophical doctrines about 
emergence. The first doctrine is that emergence is incompatible with reduction. The second is 
that emergence is supervenience; or more exactly, supervenience without reduction. I will 
claim that (with some usual meanings of these words!) both doctrines are false. This paper 
expounds the notions and analyses their general relations. But it sets aside limiting 
relations between physical theories, especially how in some cases taking a limit of some 
parameter in one theory, say $N \raw \infty$, yields another theory. Such limits {\em will} 
be a topic in the companion paper (Butterfield 2011). There the main idea will be to exhibit 
examples that combine emergence and reduction, by deducing the emergent behaviour using a 
limit of a parameter.

I should first  explain what I will mean by the contested terms `emergence', `reduction' and 
`supervenience' (Section \ref{emcenr}). Then I will give a prospectus (Section \ref{prosp}).

\subsection{Defining terms}\label{emcenr}
\subsubsection{Emergence as novel and robust behaviour; reduction as deduction; supervenience 
as determination}\label{definechoose}
I shall take emergence to mean:  properties or behaviour of a system which are {\em novel} 
and {\em robust} relative to some appropriate comparison class.  Here `novel' means something 
like: `not definable from the comparison class', and maybe `showing features (maybe striking 
ones) absent from the comparison class'. And `robust' means something like: `the same for 
various choices of, or assumptions about, the comparison class'. Often these words are made 
more precise by the fact that the system is a composite. So the idea is that its properties 
and behaviour are novel and robust compared to those of its component systems, especially its 
microscopic or even atomic components. I shall also put the idea in terms of theories, rather 
than systems: a theory describes properties or behaviour  which are  novel and robust 
relative to what is described by some other theory with which it is appropriate to 
compare---often a theory of the system's component parts.\footnote{In the companion paper, 
but not here, the idea will be made precise in terms of limits; roughly as follows. Often the 
system  is a limit of a sequence of systems, typically as some parameter (in the theory of 
the systems) goes to infinity (or some other crucial value, often zero); and its properties 
and behaviour are novel and robust compared to those of  systems described with a finite 
(respectively: non-zero) parameter. These ideas can also be put in terms of quantities and 
their values, rather than systems.}

I shall take reduction as, essentially, deduction of one theory from another; though usually 
the deduction goes through only if the other theory is augmented with appropriate definitions 
or bridge-principles linking the two theories' vocabularies. This construal will mean that I 
come close to endorsing the traditional account of Nagel (1961), despite various objections 
that have been levelled against it; (details in Section \ref{No1}).

Finally, `supervenience' is a much less contested term than the other two. It is taken by all 
to be a relation between families of properties, of {\em determination} (also called 
`implicit definability'): the extensions of all the properties in one family relative to a 
given domain of objects determine the extension of each property in the other family. 
Besides, this is widely agreed  to be a weakening of the usual notion of the second family 
being definable from the first, which is called `explicit definability'. (But Section 
\ref{Beth} will describe circumstances in which it is not a weakening, i.e. in which 
supervenience collapses to the usual notion.) Since the definitions used in a Nagelian 
reduction would be of this explicit sort, supervenience is widely taken to be a weakening of 
Nagelian reduction.

To sum up: my notion of emergence is not ``formalist'' or ``logical''; but my notions of 
reduction and supervenience {\em are}.

\subsubsection{Avoiding controversy}\label{avoidcontr}
I concede that there is already scope for vagueness, subjectivity, and thereby controversy! 
As regards emergence: What counts as an appropriate comparison class? And for a fixed class, 
being `striking' is likely to be subjective. And how many, or how varied, must the choices or 
assumptions about such a class be so as to yield robustness? Besides, philosophers and 
logicians recognize various definitions of `definability'. The variety relates both to 
metaphysical controversies about the identity of properties, and to logical contrasts between 
finite and infinitary definability (cf. Sections \ref{No1} and \ref{No2}).
These questions reflect the ongoing debate among philosophers over the correct or best 
definition of emergence, and of its apparent contrary, reduction or reducibility. 
(Silberstein (2002) is a recent survey of proposals for both emergence and reduction; cf. 
also Bedau (2003, pp. 157-161).) For example:  philosophers disagree about what counts as 
explanation; and so they disagree about what it takes for one theory (or something similar, 
such as a model) to explain or reduce another theory, or model or phenomenon.

But in this paper, these ambiguities and controversies will not be a problem, for three 
reasons. (1) {\em `Similarity'}: My meanings for the contested terms, `emergence' and 
`reduction', will be similar to those of many authors, scientific and philosophical.

(2) {\em `Clarity'}: I will not need to claim that my meanings are in some sense the ``best'' 
meanings to attach to the terms; for example by being derived from a survey of examples, or 
by satisfying a list of desiderata. In fact, I doubt that there are such best meanings. As 
regards `emergence', this is widely agreed. Many authors begin clearing the ground for their 
discussion by announcing their understanding of `emergence': usually stressing, as I do, 
novel behaviour---though  less often its robustness.

Similarly, I doubt that there is a best meaning of `reduction'. Though I will often disagree 
with objections made against Nagel's account of reduction (details in Section \ref{No1}), at 
the end of the day, I will be willing to say `what's in a name?'. That is, my heterodox or 
traditionalist endorsement of Nagel will be in part stipulative: it will be clearer, and 
convenient, to mean by `reduction' the comparatively precise relation of deducibility using 
appropriate definitions or bridge principles.

One reason why this will be clearer and convenient is the obvious one, that it makes 
supervenience a weakening of reduction ({\em modulo} the collapse in Section \ref{Beth}). 
Other reasons will be given in the preamble to Section \ref{No1}. But in short, they will be 
a matter of tactics. This paper has limited space and limited scope. So: since I {\em can} 
establish my main claims using my narrow deductive conception of reduction---but without 
settling on a general account of reduction (or of related topics like explanation, laws and 
causation), and even without assessing the pros and cons of Nagel's account---I do 
so.\footnote{But (in partisan spirit!) I take heart from recent defences of Nagel's account, 
and similar accounts such as Schaffner's (1967, 1976), against their critics; e.g. Endicott 
(1998), Marras (2002), Klein (2009), Needham (2009), Dizadji-Bahmani, Frigg and Hartmann 
(2010).\label{heart}}

Admittedly, there will be one controversial stance I {\em do} adopt. I will reject the 
multiple realizability argument, developed by authors such as Putnam and Fodor. Here I will 
endorse a refutation by Sober (1999), which I believe is decisive and unfairly neglected. But 
since the multiple realizability argument applies equally to supervenience as to reduction, I 
will postpone this topic till Section \ref{superdef}.

To sum up: the result of my limited space and scope, and of my postponement, will be that my 
Section on reduction (Section \ref{No1}) will be more concerned with formal deductions, less 
controversial (and mercifully shorter!), than you would nowadays expect for a discussion of 
reduction.

(3) {\em `Tension'}: In any case, my meanings of `emergence' and `reduction' are not chosen 
trickily, just so as to easily attain my aims. To the charge of carefully defining terms so 
as to immediately secure the desired conclusion, I plead innocent! To explain this, I need to 
state my aims. I said at the start that I aimed to show that emergence is compatible with 
reduction (usually considered its contrary), and that emergence is not the same as (what is 
often called) ``mere supervenience'', i.e. supervenience without reduction, taken as 
deduction using appropriate definitions. More positively, the two papers together aim to show 
that:---\\
\indent (i) Emergence is not in all cases failure of reduction, even in this strong sense of 
reduction. There are cases combining emergence with reduction---as well as cases of one of 
these without the other (of course);\\
\indent (ii) Nor is emergence in all cases mere supervenience (Section \ref{No2}); nor is it 
in all cases failure of mere supervenience.\\
In short: we have before us a varied landscape---emergence is {\em independent} of these 
other notions. Thus I submit that my meanings of `emergence', `reduction' and `supervenience' 
are precise and strong enough to make it worth arguing for these independence claims.

Besides, my meanings of `emergence' and `reduction' are in tension with each other: since 
logic teaches us that valid deduction gives no new ``content'', how can one ever deduce novel 
behaviour? (Of course, this tension is also shown by the fact that many authors who take 
emergence to involve novel behaviour thereby take it to also involve irreducibility.) My 
answer to this `how?'  question, i.e. my reconciliation, will lie in the use of limits. I 
have postponed that topic to the companion paper. Suffice it to say here that the idea is 
that one performs the deduction after taking a limit of some parameter: so the main morals 
will be that in such a limit there can be novelty, compared with what obtains away from the 
limit, and that ({\em pace} some authors) this should count as reduction, not irreducibility.

To sum up these three reasons, (1) to (3): my meanings of `emergence' and of `reduction' are 
similar enough to other authors' meanings, and clear enough  in themselves, and in tension 
enough with each other, to make it worth arguing that we can often ``have our cake and eat 
it'': that there are cases of emergence and reduction. And similarly for emergence's being 
independent of supervenience: there are cases of emergence without supervenience, and {\em 
vice versa}.

\subsubsection{The middle of a spectrum}\label{middle}
Note that my construal of emergence falls in the middle of a spectrum. On the one hand, it is 
more general than various specific (and sometimes speculative) scientific proposals for 
understanding higher-level or emergent theories or phenomena. In recent decades, these 
proposals have been drawn from either or both of: (i) a handful of general ideas, including 
computational intractability, non-linearity, heirarchy, and scaling; and (ii) a handful of 
proposed paradigms or schemes; e.g. cellular automata or the renormalization group or 
self-organized criticality. (Simon (1996, Chapter 7) sketches some of these.) Thus I will not 
incorporate any of these ideas and paradigms in my notion of emergence. (However, of the 
companion paper's four main examples,  one example will involve scaling, and another the 
renormalization group.)

 This exclusion is not just a matter of it being useful for philosophical discussion to keep 
the notion of emergence general; (nor just of my ignorance of most of these ideas and 
paradigms!). I also admit to a curmudgeonly scepticism about the prospects of a science of 
emergent phenomena being unlocked by some small handful of ideas. That is: I doubt that any  
such  handful can be the key to understanding all emergent phenomena.\footnote{I will not try 
to support my scepticism. Suffice it to say that I am not alone: for example, Frigg (2003) 
criticizes the claims that self-organized criticality is a universal theory.} So it is best 
to construe `emergence' generally.

But I should also make a less curmudgeonly response to the general ideas (i) and proposed 
paradigms (ii). Some philosophers  have in recent years agreed that supervenience is too weak 
to characterize emergence, and have then proposed that an idea or paradigm of kinds (i) or 
(ii) {\em does} characterize emergence; (details in Section \ref{411CisEmgceMere}). This 
prompts the following response. As I said in Section \ref{avoidcontr}, I agree that emergence 
is not supervenience, though for reasons of my own (Section \ref{No2}). And I am not 
essentialist about how to use `emergence'. So I can be pluralist and say that, for example, 
computational intractability in cellular automata, gives an acceptable sense of `emergence'. 
(In fact, this is the proposal of Bedau (1997).)

My construal of emergence is also weaker than some philosophical (as against scientific) 
proposals. Authors such as Bishop and Hendry have advocated in many papers---e.g. Bishop 
(2008, especially Sections 1 and 4.2) and Hendry (2010, especially Sections 2 and 
4)---doctrines of `top-down causation' according to which higher-level entities or properties 
`have a causal influence on the flow of events at the lower levels ... from which they 
emerge' (Kim, 1999, p. 143). I myself reject such doctrines; (and not just because of 
uneasiness about the notion of causation in general: Butterfield 2007, Section 2). But as 
mentioned in Section \ref{avoidcontr}, I can in this paper set such doctrines aside, since I 
can establish my claims without tangling with the notion of causation.

So much for how my construal of emergence is more general than various proposals. On the 
other hand, it is more specific than the idea of solving a problem by finding an appropriate 
`good' set of variables (quantities) and-or an appropriate `good' approximation
scheme. I admit that this idea is all-important: it is endemic in science, and a great deal 
of creativity (skill, imagination---and luck!) can be required to find good variables and-or 
good approximation schemes. I also admit that it is natural here to talk of `emergence': to 
say that the good variables, and-or their behaviour, `emerge'---perhaps, emerge under the 
conditions that the good approximation scheme applies. I also admit that the topic of good 
variables and-or approximation schemes is philosophically rich---and philosophers have 
addressed it (often under the headings of `idealization' and `modelling').

But it is clearly  a more general topic than novel and robust behaviour. One sees this if one 
tries to list some of the considerations  that are relevant to goodness. In short: good 
variables are often a matter of one or more of: being few in number and autonomous, i.e. 
having uncoupled equations; being easily calculated with; being suited to the given problem; 
helping insight, e.g. by being suggestive for another theory. A similar list might be drawn 
up for approximation schemes. Thus, goodness is surely a logically weaker and more 
heterogeneous concept than emergence construed as novel and robust behaviour. Presumably, the 
closest link between them is  that autonomous equations make for robustness, in the sense 
that these equations' variables have values (and over time: the equations have solutions) 
that are the same whatever the values of many other variables.

I think it a merit of my construal of emergence that it steers this middle course. Surely it 
would hardly be news if I told you that an idea so specific as cellular automata, or an idea 
so general as good variables and approximation-schemes, was compatible with reduction.

\subsection{Prospectus}\label{prosp}
As I said in Section \ref{avoidcontr}, my aim is to show that emergence is independent of 
both (a) reduction and (b) supervenience. I spell out this aim in Section \ref{ers}. In 
Section \ref{No1}, I argue for (a), the independence from reduction, though looking ahead a 
little to the companion paper's treatment of limits, and its physical examples. And this 
independence holds, with a strong understanding both of `emergence' (i.e. `novel and robust 
behaviour') and of `reduction' (viz. deduction using appropriate definitions). In Section 
\ref{No2}, I argue for (b), the independence from supervenience---having first explained and 
discussed supervenience in Section \ref{superdef}.

Of my various subsidiary claims, let me state here four main ones. Unsurprisingly in such a 
well-worked area of philosophy, each is shared with some other authors:---\\
\indent (i): I defend the traditional Nagelian conception of reduction (Section \ref{No1}); 
(cf. the authors listed in footnote \ref{heart});  \\
\indent (ii): I deny that the multiple realizability argument causes trouble for reductions, 
or ``reductionism'' (Section \ref{superdef}); (cf. Sober 1999); \\
\indent (iii): I stress the collapse of supervenience into deduction via Beth's theorem 
(Section \ref{Beth}); (cf. Hellman and Thompson 1975);\\
\indent (iv): I adapt some known examples  to show supervenience  without emergence and {\em 
vice versa}; (for supervenience  without emergence, my examples are from Hellman and Thompson 
(1977); and for the opposite, I follow several authors' citation of quantum entanglement).

So let me once more sum up my claims. Emergence is not in all cases failure of reduction, 
even in the strong sense of deduction using appropriate definitions; (Section \ref{No1}).
Nor is emergence in all cases supervenience; nor is it in all cases failure of supervenience 
(Section \ref{No2}). In short: we have before us a varied landscape---emergence is {\em 
independent} of these two notions.

\section{Emergence is independent of reduction and supervenience}\label{ers}
In this Section, I
spell out some assumptions and jargon (Section \ref{assumejargon}), and the eight failures of 
logical implication that my claims of
independence entail (Section \ref{8fail}). Of these eight, most of the interest lies in three 
cases: (a) emergence combined with reduction, (b) supervenience without emergence and (c) 
emergence without  supervenience. My defence of case (a) will depend on examples of limiting 
relations between theories which I have postponed to the companion paper. But the general 
shape of case (a) will be clear from Section \ref{No1}. Cases (b) and (c) will be covered by 
Section \ref{No2} (Sections \ref{2.4.3.A} and \ref{2.4.3.B} respectively).

\subsection{Assumptions and jargon}\label{assumejargon}
It will be clearest to take each of the three notions, emergence, reduction and 
supervenience, as a relation between {\em theories}. For the most part, I shall understand a 
`theory', as usual in logic, as a set of sentences of an interpreted language closed under 
logical consequence.

This is likely to trigger any (or all!) of three complaints. The first is specific to my 
topic; the second comes from general philosophy of science; the third from metaphysics.\\
\indent (1) {\em `Theories?'}: It is wrong to describe all cases of emergence, reduction and 
supervenience, as a relation between theories, so understood. (For a glimpse of the 
alternatives, cf. the Figures in Silberstein (2002, p. 89-90).)\\
\indent (2) {\em `Syntax?'}: Whatever one's account of emergence, reduction and 
supervenience, this understanding of theories, viz. the traditional syntactic view, has been 
refuted in favour of the semantic view that a scientific theory is a set of models. \\
\indent (3) {\em `Properties?'}: In metaphysics, supervenience is almost always defined as a 
relation between two sets of properties, not as a relation between theories. So: what 
justifies my change?

My reply is, in short, that I reject (2) and (3), and this prompts me to ignore (1). There 
are three main points to make against (2) and (3); the second and third will become clearer 
in Section \ref{No2}.\\
\indent (i): The syntactic view is more flexible than is often admitted. The theory's 
language need not be formalized, nor even finitary; nor need the underlying logic be 
first-order. I contend that with this liberal construal, the syntactic view can describe 
perfectly well the phenomena in scientific theorizing that advocates of the semantic view 
tout as the merits of models. Indeed, this is hardly surprising. For to present a theory as a 
set of models, you have to use language, usually (at least in part) by saying what is true in 
the models; and on almost any conception of model, a model makes true any logical consequence 
of whatever it makes true.  \\
\indent (ii): I said that I understood `theory' syntactically `{\em for the most part}'! For 
in Section \ref{No2} I will need the idea of a class of models that is {\em not} the set of 
models of a given (syntactic, indeed first-order) theory. And I need this idea, not just for 
one point among many, but for a crucial point about supervenience that the literature about 
supervenience has woefully neglected. This leads to (iii).   \\
\indent (iii): Turning to (3), there is less dispute here than might appear. In short, the 
supervenience literature distinguishes {\em local} and {\em global} supervenience, and we 
will see reason to prefer the latter notion---which is tantamount to my tactic of taking 
supervenience as a relation between theories. Details in Section \ref{objsoglobal}.\\
\indent These points, (i) to (iii), suffice to answer (2) and (3). But I agree that they do 
not refute (1), especially as regards emergence. I must concede that perhaps some cases of 
emergence, or reduction or supervenience (even in my senses), are not happily described as a 
relation between theories, even liberally construed. I can only say that I know of no such 
cases, but hope that my claims could be carried over to them.

So I now propose to argue that these three relations between theories, emergence, reduction 
and supervenience, are mutually independent. I will confine myself to arguing for {\em 
logical} independence, i.e. for these relations not logically implying each other (and 
likewise, their negations). Agreed, you might complain that this is a weaker and easier aim 
than arguing for some sort of nomological independence. For even if I show that a logical 
implication fails, e.g. from emergence to non-reduction (by showing a case of emergence and 
reduction), you might reply that some weaker, e.g. nomological, implication
might yet hold. That is, you might say that emergence subject to some realistic constraints, 
e.g. subject to some
laws, excludes reduction, i.e. implies non-reduction. Fair comment. But in the debate about 
emergence ``vs.'' reduction, with its several different usages and controversies, we will 
have enough work to do, just to establish logical independence. Consider the open sea of 
considerations bearing on how you should choose your ``realistic constraints'': sufficient  
unto the day is the work thereof!

I turn to the jargon about the relations of reduction and supervenience.
The traditional philosophical jargon is that one theory $T_1$ {\em reduces} to, or {\em is 
reducible to}, another $T_2$ if, roughly speaking, $T_1$ can be shown to be part of $T_2$. 
And the Nagelian tradition treats this as $T_1$  being derived from $T_2$, subject perhaps to 
further constraints, e.g. about the derivation counting an explanation of its conclusion; 
(more details in Section \ref{No1}).

The first thing
to say is that we must beware of the conflicting, indeed converse, jargon common in physics! 
The physics jargon is that in such a case, it is
$T_2$ that reduces to  $T_1$ (typically in some limit of some parameter of $T_2$). The idea 
is of course that `reduces to' means `simplifies to': for example, physicists say that  
special relativity reduces to Newtonian mechanics as the speed of limit $c$ tends to 
infinity.\footnote{Among philosophers, Nickles (1973) has stressed the contrasting jargons. 
He also argues that the physicists' notion, which he calls `reduction$_2$', is not merely the 
converse of the philosophers' notion (`reduction$_1$'), i.e. essentially the Nagelian notion 
of derivation subject to further constraints. While reduction$_1$ is `the achievement of 
postulational and ontological economy and {\em is} obtained chiefly by derivational reduction 
as described by Nagel ...[reduction$_2$ is] a varied collection of intertheoretic relations 
rather than a single distinctive logical or mathematical relation [whose importance] lies in 
its heuristic and justificatory roles' (p. 181). I agree: and I also believe (although I will 
not argue it here) that a good deal of the `New Wave' critique of Nagel wrongly construes him 
as aiming to give an account of reduction$_2$; cf. footnote \ref{heart}.}

  I will adopt the philosophers' jargon. But to avoid confusion between $T_1$ and $T_2$, I 
will use mnemonic subscripts. For the reducing theory, I will use $b$: `b' is for 
bottom/basic/best; `bottom' and `basic' connoting microscopic and fundamental, and `best' 
connoting a successor theory. And  for the reduced theory, I will use $t$: `t' is for 
top/tangible/tainted; `top' and `tangible' connoting macroscopic and observable, and 
`tainted' connoting a predecessor theory. So to sum up: I will follow the philosopher, saying 
that a
theory $T_b \equiv T_{\rm{bottom/basic/best}}$ reduces $T_t \equiv 
T_{\rm{top/tangible/tainted}}$; and that $T_t$ reduces to, or is reducible to, $T_b$;
whereas a physicist would  say that
$T_b$ reduces to $T_t$ (typically in some limit of a parameter).
Thus the broad picture is that $T_t$ is reduced to $T_b$ by: \\
\indent (i) being shown to be  logically deducible from $T_b$, usually together with some 
judiciously chosen definitions; (details in Section \ref{defext}); and maybe also \\
\indent (ii) satisfying some further constraints e.g. about explanation; (details in Section 
\ref{philcritdefext}).

Once the relation of deducibility, in (i) above, is made precise, there is an apparent 
weakening of it, which in logic is called {\em implicit definability} and in philosophy is 
called {\em supervenience} (or {\em determination}). Roughly speaking, it is deducibility, 
with the allowance of definitions that are infinitely(!) long. But it turns out to be a 
delicate matter whether this allowance really secures any extra generality, i.e. whether 
supervenience is in fact weaker (more general) than (i)'s deducibility. This point is 
well-known in logic (under the name `Beth's theorem', proven in 1953); and will support my 
reconciliation of emergence and reduction. But it has been woefully neglected in the 
philosophical literature about supervenience.  More details in Section \ref{precise2}.

Finally, concerning jargon:
 taking emergence to be a relation between theories, I will say that $T_t$ is emergent from 
$T_b$. And `logical independence' will mean `failure of all implications'. More precisely, 
propositions $p$ and $q$ are logically independent if all four truth-combinations are 
possible. That is: $p$ does not imply $q$, so that $p \& \neg q$ is possible; nor does $\neg 
p$ imply $q$, so that $\neg p \& \neg q$ is possible; and so on for the other two cases. (In 
yet another jargon: the distinctions, $p$ vs. $\neg p$, and $q$ vs. $\neg q$, cut across one 
another.)

\subsection{Eight implications fail}\label{8fail}
 Combining these bits of jargon: for emergence to imply reduction would require that for all 
theories $T_t, T_b$, if $T_t$ is emergent from $T_b$, then $T_t$ is reduced to $T_b$; and so 
on for the other three putative implications relating emergence and reduction; and for the 
four putative implications relating emergence and supervenience.

Thus I claim that all eight implications fail. This claim is less complex, and less 
contentious, than it might first appear, for two reasons; (and not just because logical 
independence is weak, compared with some sort of nomological independence!).

First: while I take emergence as an informal concept (viz. novel and robust behaviour; 
Section \ref{emcenr}), I take reduction, and so also its weakening, supervenience, as a 
formal concept, viz. deducibility. That is: I will abjure (for reasons given in Section 
\ref{2.2.2.A}) the further constraints mentioned in (ii) of Section \ref{assumejargon}. This 
informal-formal contrast makes it easier to break the implications.

Second: some of the eight
cases are very straightforward, either because (i) they are obvious, or because (ii) they 
follow from other cases; as follows. Let us write $E, R$ and $S$ as an obvious notation.

\indent As to (i): Of course, emergence and reduction are usually regarded as mutually 
exclusive, so cases of $R \& \neg E$ (i.e. two theories where $T_t$ is reduced to $T_b$, but 
$T_t$ is not emergent from $T_b$), and $E \& \neg R$ are unsurprising. Besides, theories 
about unrelated topics will provide cases of $\neg E \& \neg R$. So for emergence and 
reduction, the only surprising case I need to establish is: $E \& R$, i.e. two theories where 
$T_t$ is reduced to, but also emergent from, $T_b$. I address this in Section \ref{No1}; 
though as announced, some of my defence is postponed to the companion paper, with its 
examples from physics.

\indent As to (ii): In general, supervenience is a weakening of reduction: if $T_t$ is 
reduced to $T_b$, then $T_t$ supervenes on $T_b$. So a case of $E \& R$ is {\em ipso facto} a 
case of $E \& S$, and thus shows that emergence does not exclude supervenience. Also: if 
emergence (or its negation) implied reduction, it would imply supervenience; and 
(equivalently but more relevantly), if emergence (or its negation) does not imply 
supervenience, it {\em a fortiori} cannot imply reduction. And I shall argue that indeed 
emergence does not imply supervenience; (nor of course does its negation).

This possibility, $E \& \neg S$, is worth emphasising, and not just because it implies that 
$E \& \neg R$ is also possible. And similarly, the other possibility, $\neg E \& S$, is worth 
emphasising. For each of them refutes a widespread  philosophical proposal about emergence: 
viz. that
emergence is precisely ``mere supervenience'', i.e. supervenience that is not reduction. 
Agreed: since `emergence' is vague, and philosophers are at liberty to tie it down as they 
see fit, this doctrine is by no means a consensus. But it is at least worth showing that for 
my meaning of emergence, as novel and robust behaviour, it fails. Details will be in Section 
\ref{SEindpdt}.  For the moment, I just note that since theories about two unrelated topics 
will provide cases of $\neg E \& \neg S$, and cases of $E \& R$ also establish $E \& S$, it 
follows that for emergence and supervenience, I only need to establish two cases: $E \& \neg 
S$, and $ \neg E \& S$.

It follows from (i) and (ii) that the three cases I need to establish are: $E \& R$, $E \& 
\neg S$, and $\neg E \& S$. As I mentioned, an idea common to all the cases will be that 
novelty and robustness are informal, non-logical, ideas, while I take reduction and 
supervenience as formal ideas: so it is less surprising that the distinction, emergence vs. 
non-emergence, cuts across reduction vs. non-reduction, and across supervenience vs. 
non-supervenience.

\section{Emergence vs. reduction? No!}\label{No1}
The main claim of my `No!' is that a theory $T_t$ can describe novel and robust behaviour, 
while being reduced to an appropriate theory $T_b$: that there are cases of $E \& R$. Most of 
this Section is devoted to this. This will lead at the end of the Section to $E \& \neg R$.

Again I admit that I will emphasise a logical, and so ``cut and dried'', conception of 
reduction as deduction using judiciously chosen adjoined definitions. This will make my $E \& 
R$ claim less contentious than it might seem. For (as I mentioned in Section 
\ref{assumejargon})  philosophers often take reduction to require more than deduction. They 
propose further constraints e.g. about explanation, which in some cases amount to a 
prohibition of novel behaviour, so that reduction indeed excludes emergence in my 
sense.\footnote{On the other hand, definitions of supervenience almost never include a 
prohibition of novelty, so that it will be easier to argue later for the  case $E \& S$,  
than here for $E \& R$.} (With such a prohibition in force, the issue of robustness does not 
arise.)

 Nevertheless it is worth articulating cases of $E \& R$, where $R$ represents a deductive 
conception of reduction, for three reasons. The first two are easy to state. The third arises 
from an objection to the second, and will be a matter of limiting this paper's scope, and of 
dividing the philosophical labour between this Section and Section \ref{superdef}.

First: articulating these cases shows that the power of reduction, i.e. deduction, is 
stronger than commonly believed. Second: it shows that philosophers' concerns about which 
further constraints reduction should satisfy tend to put the emphasis in the wrong place. It 
is after all just a matter of words whether to call a deduction of novel behaviour a 
reduction (because of the deduction) or not (because of the novelty). What matters, 
scientifically and indeed conceptually, is that you have made the deduction: your brick for 
the rising edifice of unified science!

A philosopher might object to this second reason that it equivocates---and that one 
disambiguation is false and anti-philosophical. Agreed, one can define words as one sees fit, 
and a deduction of novel behaviour is a scientific achievement. But that in no way shows that 
philosophers' concerns about further constraints on reduction are mis-directed. For one of 
the tasks of philosophy of science is to assess how well integrated our theories 
are.\footnote{For a persuasive and detailed case that this is the main task of philosophy of 
science, cf. Ladyman and Ross et al. (2007, pp. 27-53).} Indeed: are they integrated enough 
in terms of notions like explanation and the identity of theoretical entities or properties, 
that taken together they merit the metaphor `rising edifice', rather than `shambolic 
patchwork'?! However one uses `reduction' and other words, philosophers should assess how 
deductive connections between theories relate to notions like explanation and the identity of 
entities or properties.

I agree. But I will nevertheless emphasise reduction as deduction, and downplay these 
notions, for two reasons. First, as I said in (2) of Section \ref{avoidcontr}: this tactic is 
partly a matter of this paper's limited space, and limited scope. These notions, and their 
role (if any) in reduction, are beset by controversies I have no space to address. Besides, 
my scope is limited. In particular, I do not urge that all known examples of emergence are 
also examples of reduction; and I do not need to take a position in these controversies, nor 
to settle on a general account of reduction. (But as mentioned in footnote \ref{heart}, I 
favour the Nagel-Schaffner account, not its `New Wave' critics.)

Second: this tactic is partly a mere matter of dividing labour between this Section and 
Section \ref{superdef}. It will be clearer to first treat reduction in logical terms, and to 
postpone these controversial notions till we discuss supervenience in Section \ref{superdef}. 
That is hardly surprising: since (as I mentioned) supervenience is a weakening of reduction's 
core notion of deducibility, these notions also loom large in discussing supervenience. 
Besides, we will see this connection to supervenience already in Section \ref{2.2.2.A}, where 
we return to the further constraints that might be required of reduction.

So I turn to details about reduction as deduction (Section \ref{defext}). These details will 
lead us back to the further constraints (Section \ref{philcritdefext}).

\subsection{Definitional extensions}\label{defext}
\subsubsection{Definition and comments}\label{defextdefined}
Recall the basic idea that one theory $T_t$ is reduced to
another $T_b$ by being shown to be a part
of $T_b$. The notion of {\em definitional extension} makes this
idea precise. The
syntactic conception of theories immediately gives a
notion of $T_t$ being a part of $T_b$, viz.\ when $T_t$'s
theorems are a subset of the $T_b$'s. (This is called $T_t$
being a sub-theory of $T_b$.) However, one needs to avoid
confusion that can arise from the same predicate (or other
non-logical symbol) occurring in both theories, but with
different intended interpretations. This is usually
addressed by taking the theories to have disjoint
non-logical vocabularies. Then one defines $T_t$ to be a
{\em definitional extension\/} of $T_b$, iff one can add to
$T_b$ a set $D$ of definitions, one for each of $T_t$'s non-logical symbols,
in such a way that $T_t$ becomes a sub-theory of the augmented theory $T_b \cup D$.
That is: In the augmented theory, we
can prove every theorem of $T_t$. This is the core idea of Nagelian reduction, with the 
definitions being called `bridge laws' (or `bridge principles' or `correspondence 
rules').\footnote{The standard references are Nagel (1961, pp. 351-363; and 1979); cf. also 
Hempel (1966, Chapter 8). Discussions of bridge laws are at Nagel (1961, pp. 354-358; 1979, 
366-368) and Hempel (1966, pp. 72-75, 102-105). Section \ref{philcritdefext} will mention 
revisions by other authors such as Schaffner.}

This prompts five logical comments, in roughly ascending order of specificity. For most of 
these comments, the theory's language need not be formalized, nor even finitary; nor need the 
underlying logic be first-order. But we shall have to be more specific when we discuss 
supervenience in Section \ref{No2}, especially the conditions under which it collapses  in to 
definitional extension; cf. Section \ref{Beth}. And as I explained in this Section's 
preamble, some controversial notions such as explanation and the identity of properties will 
be for the most part postponed to Sections \ref{philcritdefext} and \ref{superdef}. For 
example, I will take a definition of a predicate $P$ to be a statement that $P$ is 
co-extensive with an open sentence (cf. (3) below). I will not here consider whether the open 
sentence must be short and-or conceptually unified; nor will I consider over how large a set 
of interpretations of the language (``possible worlds'') the co-extension must hold. But of 
course I agree that these issues of how brief or unified, and of how modally strong, such a 
definition needs to be are central to discussions of reduction, explanation and the identity 
of properties.

(1): {\em Choosing definitions}: The definitions must of
course be judiciously chosen, with a view to securing the
theorems of $T_t$. And it can take a lot of creativity (skill, imagination---and luck!) to 
frame such definitions: recall from Section \ref{emcenr} the creativity in finding good 
variables and good approximation schemes.

(2): {\em Allowing long definitions and deductions}: There is no requirement that the 
definitions of $T_t$'s terms, or deductions of its theorems, be short. A definition or 
deduction might be a million pages long, and never formulated by us slow-witted humans. So 
when in this Section and my later examples (e.g. Section \ref{defextexamples}), I praise the power of deduction, I am in part 
celebrating how brief, and so how humanly comprehensible, many such definitions and 
deductions are. For informal yet mathematized theories, this brevity is sometimes won by 
having compact and powerful notation; the classic example is Leibniz's $dy / dx$ notation for 
the calculus. For formalized theories the brevity is sometimes won by giving $T_b$ a rich set 
of operations for constructing definitions (e.g. including limit operations) and a 
correspondingly strong underlying logic to sustain deductions. This is illustrated by (3) and 
(4).

(3): {\em Definition as co-extension}: The core idea of a definition is that it is a 
statement, for a predicate, of co-extension; and for a singular term, of co-reference. Thus a 
definition of a primitive predicate $P$ of $T_t$ will be a universally quantified 
biconditional with $P$ on the left hand side stating that $P$ is co-extensive with a right 
hand side that is a compound predicate (open sentence) $\phi$ of $T_b$ built using such 
operations as Boolean connectives and quantifiers. Thus if $P$ is $n$-place, the definition 
is:  $(\forall x_1)...(\forall x_n)(P(x_1,...,x_n) \equiv \phi(x_1,...,x_n))$. This scheme 
can be extended in logic textbooks' usual way to give definitions of individual constants and 
function symbols, viz. by analysing them in terms of predicates, using Russellian 
descriptions.\\
\indent Agreed, this core idea is limited in three respects, which have made some philosophers doubt that it is useful for understanding reduction in science. I take up two of these respects in (4) and (5). The third, I mentioned just before (1): to secure the identity of properties, we surely need not just actual co-extension, but co-extension in a large set of ``possible worlds''. I take this up in Sections \ref{philcritdefext} and \ref{superdef}.

(4): {\em Getting new objects}: But the scheme in (3) does not extend the domain of 
quantification, while the objects (systems) described by $T_t$ often seem distinct from those 
of $T_b$. As I see matters, this is addressed by three familiar tactics, which are often 
combined together, and which we can dub, in order: `theoretical identification', `logical 
construction' and `curly brackets'.\\
\indent {\em Theoretical identification}: What seems so might, surprisingly, not be so. The 
reduction might identify objects of $T_t$ with some of $T_b$. A standard example is Maxwell's 
theoretical identification of light with electromagnetic waves; (e.g. Sklar 1967, p. 118; 
Nagel 1979, p. 368).\\
\indent {\em Logical construction}: We might be willing to give up what seems so. That is: we 
might accept the reduction's offer of logical constructions in $T_b$ as replacements for the 
objects in $T_t$. To give a convincing example, one of course has to imagine people's views 
before they were persuaded of the replacement. For example: imagine Kronecker saying that God 
gave us the integers, and then accepting Frege's or Russell's reduction as a logical 
construction or replacement of them. This leads to the third tactic.\\
\indent {\em Curly brackets}: Set theory provides a famously powerful way to define (should 
we say create or discover?!) objects with a formal structure appropriate for well-nigh any 
theoretical role in a mathematical or scientific theory.

(5): {\em Functional definitions and multiple realizability}: Definitional extensions, and 
thereby Nagelian reduction, can perfectly well accommodate what philosophers call `functional 
definitions'. These are a kind of definition of a predicate or other non-logical symbol (or 
in ontic, rather than linguistic, jargon: of a property, relation etc.) that are {\em 
second-order}, i.e. that quantify over a given base-set of properties and relations. The idea 
is best explained by the philosophically familiar example of functionalism in the philosophy 
of mind.\\
\indent Functionalism holds that each mental property (or relation), e.g. the property of 
being in pain, can be defined by its place in a certain pattern of relations (typically, 
causal and-or lawlike relations) between various physical properties and-or their instances. 
So a first guess for a functional definition of `being in pain' might be: `the property of an 
animal that is typically caused by damage to its tissues and typically causes aversive 
behaviour'. But functionalists emphasize that this {\em definiens} is likely to have 
different referents in different varieties of animal; and all the more likely, for varieties 
with disparate nervous systems, e.g. a mollusc, an octopus and a human. So they propose that 
`being in pain', as a predicate expressing the common property, should be defined by 
quantifying over these different referents. Adjusting our first guess, this would give the 
{\em definiens}: `the property of an animal of having some physical property  that is 
typically caused by damage to its tissues and typically causes aversive behaviour'.\\
\indent Some jargon: the definition is called a `functional definition', and pain a 
`functional property' because it is not only second-order but involves a pattern of causal 
and lawlike relations; such a pattern is called a `functional role'; the different physical 
properties in the different varieties of animal are called `realizers' or `realizations' of 
pain; and the fact that there are various such properties is called `multiple 
realizability'.\\
\indent So much by way of explaining functional definitions and their associated jargon. It 
remains to make three points about them.\footnote{Since my topic is reduction, not 
functionalism about the mind, I set aside philosophers' objections to the latter. But note 
that apart from the widespread view that the qualitative ``feel'' of pain prevents a 
functional definition of pain, Field makes another objection about mental representation 
(1978, Sections 2,3, pp. 24-40). For brevity, I also set aside the question whether the 
functional definition is obtained from conceptual analysis or from the contingent claims of a 
scientific theory. Though my examples of definitions of pain suggest the former (and so: 
analytical functionalism in the philosophy of mind), the latter occurs very often: as is 
clear from the non-mental examples in points (ii) and (iii) below, and as is often remarked 
in the literature (e.g. Needham 2009, p. 105, p. 107).}

\indent  (i): It is evident that a definitional extension can incorporate functional 
definitions, for two reasons. First, as I said at the start of this Subsection, the 
underlying logic need not be first-order; so if you wish you can reflect the second-order 
status of the {\em definiendum} by using a second-order logic. Secondly, you can instead keep 
to a first-order logic by treating properties as values of first-order variables, and (very 
probably) also using the powerful apparatus of set theory (cf. `Curly brackets' in (4) 
above). For example, one of the definitive expositions of functionalism (applied not just to 
mind) adopts this latter strategy (Lewis 1970, p. 80; also e.g. Loar 1981, pp. 46-56).

\indent (ii): I have so far talked about two ``levels'': a base-set of properties and 
relations, and second-order properties and relations---in my example, the physical  and the 
mental. But we should no doubt distinguish many such levels: both as regards the mind-matter 
relation, and in other examples, such as the relation of physics to chemistry, and even 
within physics, such as the relation of nuclear physics to atomic physics, or molecular 
physics to mesoscopic physics. And between most, or even all, pairs of adjacent levels, there 
will no doubt be cases of multiple realizability. So for a theory $T_t$ at a higher level to 
be shown to be a definitional extension of one at a lower level will require a succession of 
functional definitions. As a result, the class of objects (in our example: animals) across 
which there is a  common realizer in the sense of the bottom-level taxonomy of properties, is 
likely to be small. In general, all we can expect is that this class will be the intersection 
of the various varieties defined at the various levels. Returning to (i)'s concern with a 
formal approach's choice of logic: to cope with a succession of functional definitions, and 
avoid logics of correspondingly higher order, one would no doubt adopt (i)'s second strategy, 
i.e. first-order logic plus set-theory.

\indent (iii): Some philosophers think that multiple realizability provides an argument 
against reduction. The leading idea is that the {\em definiens} of a multiply realizable 
property shows it to be too ``disjunctive'' to be suitable for scientific explanation, or to 
enter into laws. And some philosophers think that multiple realizability prompts a 
non-Nagelian account of reduction; even suggesting, {\em pace} (i), that definitional 
extensions cannot incorporate functional definitions. Both these lines of thought have 
adherents in philosophy  of mind and in philosophy of physics. For example, in philosophy of 
mind: Kim calls Nagelian reduction `philosophically empty' and `irrelevant', because Nagelian 
bridge laws are `brute unexplained primitives'; (1999, p. 134; 2006, p. 552; and similarly in 
other works: 1998, pp. 90-97; 2005, pp. 99-100). Compare also Causey's requirement that the 
bridge-laws express attribute-identities (1972, p. 412-421). Kim's own account, called a 
`functional model of reduction', takes reduction to include (a) functional definitions of the 
higher-level properties $P$ etc. and (b) a lower-level description of the (variety-specific) 
realizers of $P$ etc., and of how they fulfill the functional roles spelt out in (a). An 
example in philosophy of physics: Batterman agrees that multiple realizability spells trouble 
for Nagel, and that Kim's model is not Nagelian; and he applauds it, though with a 
reservation about how to explain features common across different varieties of realizer; 
(2002, pp. 65, 67, 70, 71-74 respectively).

\indent As (i) and (ii) suggest, I wholly reject both these lines of thought. Multiple 
realizability gives no argument against definitional extension; nor even against stronger 
notions of reduction like Nagel's, that add further constraints additional to deducibility, 
e.g. about explanation. That is: I believe that Nagelian reduction, even including such 
constraints, is entirely compatible with multiple realizability. This was shown very 
persuasively by Sober (1999). And for critiques of Kim in particular, cf. Marras (2002, pp. 
235-237, 240-247), Rueger (2006, pp. 338-342) and Needham (2009, pp. 104-108).  But as 
announced in this Section's preamble, I postpone this rebuttal till I discuss supervenience 
in Section \ref{superdef}, specifically Section \ref{411Amra}. For the controversy applies 
equally to supervenience as to reduction. Here let me just sum up by saying that I see no 
tension between Nagelian reduction and the fact of multiple realizability, with its 
consequent need for functional definitions.

\subsubsection{Examples and morals}\label{defextexamples}
So much by way of logical comments about definitional extensions. For us, the important point 
is that there are many examples of definitional
extensions in physics, especially if in line with (2)-(4) of Section \ref{defextdefined}, we give $T_b$ a suitably strong logic and-or set theory so as to ``maximize its deductive reach''. Besides, the definitions and deductions that secure the definitional extension are remarkably brief and comprehensible.

Examples arise in almost any
case where there is a deduction of one physical theory
from another. And there are many such. Agreed, `physical theory' is sometimes taken so 
broadly, e.g. `wave optics' or `classical thermodynamics', that theories are so complex and 
open-ended (and inter-related) as to be hardly formalized and hardly deducible one from 
another. But examples abound once one takes `theory' to be specific enough.

And here, `enough' need not mean `very'! There are striking examples of one {\em general} 
theory being deduced from another. Consider equilibrium classical statistical
mechanics for a strictly isolated system, with the postulate of
the microcanonical measure on the energy hypersurface. It is a
definitional extension of the classical mechanics of the
microscopic constituents, once we use an underlying logic strong
enough to express calculus of many
variables. (For the microcanonical measure is explicitly definable from the phase space's 
Lebesque measure and the gradient of the Hamiltonian.)

This example also illustrates two morals which will be important in what follows. The first 
is that claims of deducibility are of course sensitive to exactly which theories are being 
considered. Thus I concede that: (i) equilibrium classical statistical
mechanics is sometimes taken to include the ergodic hypothesis, roughly to the effect that 
the system's state gets arbitrarily close to any point in the energy hypersurface, since this 
seems necessary for justifying phase-averaging; and (ii) notoriously, the ergodic hypothesis 
is not deducible from the general microscopic mechanics---though it does follow if we add 
various more specific assumptions. So it is a case of swings and roundabouts: we just have to 
define precisely our $T_b$ and $T_t$. This point is very simple, but it will recur: first in 
Section \ref{philcritdefext}, then in Section \ref{SEindpdt},  and often in my examples in 
the companion paper.

The second moral is more positive. This example illustrates two successes of `reduction as 
deduction' which are general: indeed, endemic to both classical and quantum physics. Each is 
a general and surprising simplicity about how to describe composites in terms of their 
components, and each will be prominent in the companion paper's examples. The first is about 
states, the second about forces:---\\
\indent [1]: The  rules for defining a composite system's state-space and its quantities are 
very {\em uniform}. In particular, for state-spaces we use: Cartesian products in classical 
physics; tensor products in quantum theory).\\
\indent  [2]:  There are no fundamental forces that come into play only when the number of 
bodies (or particles, or degrees of freedom) exceeds some number, or when the bodies, 
particles etc. enter into certain configurations, or more generally, states. To put the point 
more positively: both quantum and classical physics postulate only two-body forces.

In the practice of physics, [1] and [2] are so widespread and familiar that we tend to take 
them for granted. Certainly they are hardly mentioned in the philosophy of science's 
literature on reduction.\footnote{Agreed, [2] is discussed in connection with British 
emergentism, especially in the philosophy of mind literature. Thus McLaughlin (1992) 
describes Broad's conjecture that the explanation of chemical compounds' properties would 
need to deny [2], i.e. would need to postulate what Broad called `configurational forces' 
between atoms that are exerted only when enough of them are close enough to each other. As 
McLaughlin says, this was a reasonable conjecture. He goes on to claim that it was refuted by 
the development of quantum chemistry; but this claim has been disputed, e.g. by Scerri (2007, 
pp. 920-925). Point [1], on the other hand,  seems to go un-mentioned in the philosophical 
literature.\label{Broadn}} But they are both contingent, and pieces of amazing good fortune 
for physical enquiry. They will also be important for my argument that emergence is 
independent of supervenience; cf. Section \ref{2.4.3.B}.

To sum up this eulogy for reduction as deduction: Although deducing real-life theories from 
one another is hard (cf. comment (1)), there have been some outstanding successes in physics, 
especially after 1850, when the development of many rich physical theories could be coupled 
to the powerful tools of modern logic and set theory.

A final comment: these successes also reflect the extraordinary unity of nature. I will not 
try to make this precise. But think, for example, of quantum theory's explanation of chemical 
bonding, as in footnote \ref{Broadn}; or the cosmic applicability of the periodic table 
discovered by terrestrial chemistry; or the applicability in so many kinds of dynamics of 
least-action principles. This unity is, so far as we can tell, contingent; but it is very 
striking and, for human enquiry, fortunate. Indeed, we are nowadays sufficiently confident of 
such unity that research programmes aiming to reveal disunity are often heterodox. 

But by `heterodox' I do not mean to suggest `crankish' or `incredible'. Here are two recent examples 
which, involving the high talents of a Nobel prize-winner, have the salutary effect of 
showing something my eulogy has ignored: the scientific (rather than philosophical) 
importance of establishing a {\em failure} of reduction. Thus consider:\\
\indent (i) Prigogine's programme to find novel time-irreversible laws in order to better 
explain irreversible phenomena in thermal physics;\\
\indent  (ii) Leggett's prize-winning work on superfluid helium was motivated by his 
expecting that the phenomena could not be explained by orthodox quantum mechanics---he 
believed it would need to be modified by admitting new forces, i.e. a quantum analogue of 
Broad's configurational forces, as in footnote \ref{Broadn}.\\
\indent Agreed: in example (i), the jury is still out, and most of its members are sceptical; 
while in example (ii), orthodox quantum mechanics {\em could} explain the phenomena. So these 
are indeed examples of heterodoxy; and so they are testimony to how widespread is reduction, 
and to the unity of nature.

\subsection{Philosophical criticisms of reduction as deduction}\label{philcritdefext}
Agreed, Section \ref{defext}'s  eulogy for reduction as deduction did not mention deducing 
novelty and robustness. So it does not show the compatibility of emergence and reduction, my 
case $E \& R$. Completing that task must wait for the examples in the companion paper (2011). But an 
important part of the task is now just around the corner. For---changing the metaphor---it is 
just the other side of the coin, for a point I already made in Section \ref{assumejargon} and 
the preamble to Section \ref{No1}: viz. philosophers' tendency to require reduction to obey 
further constraints, e.g. about explanation or property-identity, apart from deduction. As we 
shall see, Nagel himself concurred.

\subsubsection{Deducibility is too weak}\label{2.2.2.A}
Thus philosophers have said that even if $T_t$ is
a definitional extension of $T_b$, there can be non-formal aspects of $T_t$ that are
not encompassed by (are not part of) the corresponding
aspects of $T_b$; and that, at least in some cases, these aspects
seem essential to $T_t$'s functioning as a scientific theory. Of course, philosophers 
disagree about exactly what these aspects are! Some stress metaphysics, some epistemology.   
Thus one metaphysical aspect concerns the identity of properties. Even though every primitive 
predicate of $T_t$ is co-extensive with a compound predicate of $T_b$ (cf. (3) of Section 
\ref{defextdefined}), maybe, at least in some cases, $T_t$'s properties are different from 
those of $T_b$,
including its compound properties. And one epistemological aspect concerns explanation: maybe  
$T_t$ provides explanations that are not encompassed by $T_b$. And similarly for other 
epistemological aspects, e.g. heuristics for
modelling.

This tendency is long-established, and endorsed by the best authorities.
Thus Nagel himself (1961, pp. 358-363) adds to the
core idea of definitional extension some informal
conditions, mainly motivated by the idea that the reducing
theory $T_b$ should explain the reduced theory $T_t$; and following
Hempel, he conceives explanation in deductive-nomological
terms. Thus he says, in effect, that $T_b$ reduces $T_t$
iff: \\
\indent (i): $T_t$ is a definitional extension of $T_b$; and \\
\indent (ii): In each of the definitions of $T_t$'s terms, the {\em definiens} in the 
language of $T_b$ must play
a role in $T_b$; so it cannot be, for example, a
heterogeneous disjunction.

Some philosophers have objected that Nagel's proposal does not secure what was needed: that 
the definitions (bridge laws) should be property-identities and-or should be explained. (Cf. 
the references to Kim and Causey in (5)(iii), at the end of Section \ref{defextdefined}.) 
Other philosophers are much closer to Nagel; e.g. Sklar (1967, pp. 119-123), Nickles (1973, 
pp. 190-194).

I incline to the latter view---but I do not need to assess this debate.
For my purposes, the main point is this: although I have presented reduction as deduction, I 
can perfectly well concur with this tendency to add further contraints. As I said in the 
preamble of Section \ref{No1}, the apparent conflict is in part just a matter of words. That 
is: one can interpret `reduction' as one theory being part of (or encompassed by) another in 
a strong sense. Then there will be fewer reductions. And in particular, a reduced $T_t$ will 
{\em not} exhibit novelty relative to $T_b$, and so reduction will exclude emergence in my 
sense.  But on the other hand, one can---I do---interpret `reduction' in a weaker sense, 
requiring only deduction (as spelt out in (1)-(5) of Section \ref{defextdefined}). Then there 
will be more reductions; and the door is open to cases of reduction and novelty, even 
reduction and  emergence (i.e. both novelty and robustness).  Indeed, the door is wide open. 
For the very same considerations, e.g.
about the deduced theory $T_t$ having novel properties (or to take the other example: about 
$T_t$'s
autonomous explanations) that prompt the tradition to conclude `Here is no reduction', prompt 
someone like me to say `Here is reduction, i.e. deduction, with novelty or autonomy 
etc.'.\footnote{Agreed, the conflict is {\em only} in part a matter of words. Again, I agree 
that there are philosophical issues, which the preamble of Section \ref{No1} postponed to 
Section \ref{superdef}, about how such deductions relate to explanation etc; cf. especially 
Section \ref{superinf}.}

So much, for the moment, by way of arguing for emergence with reduction, for my case $E \& 
R$. To complete the argument, I would have to say more about novelty; and to discuss 
robustness, which this Section has ignored. But I postpone that until the companion paper's 
examples, which will show how one can get novelty and robustness by taking a limit.

\subsubsection{Deducibility is too strong}\label{2.2.2.B}
I end this Section by turning to another traditional criticism of reduction  as deduction: 
not (as above) that it is too weak and further constraints need to be imposed, but that it is 
too strong. Not only do I feel duty-bound by tradition to report this criticism. Also, since 
it weakens the concept of reduction, it makes cases of emergence and reduction easier to come 
by---grist to my mill.

The criticism is that definitional extension, and thereby proposals like
Nagel's that incorporate it (and any variants that keeps his
clause (i), only adjusting his (ii)), is too {\em
strong} for reduction. (This objection is made by e.g. Kemeny and Oppenheim (1956), and 
Feyerabend (1962).) For in many cases where $T_b$ reduces $T_t$, $T_b$ {\em corrects}, rather 
than implies, $T_t$.  One standard example is Newtonian gravitation theory ($T_b$) and 
Galileo's
law of free fall ($T_t$). This $T_t$ says that bodies near the earth fall with constant 
acceleration. This $T_b$ says that as they fall, their acceleration increases, albeit by a 
tiny amount. But surely $T_b$ reduces $T_t$. And similarly in many famous and familiar 
examples of reduction in physics: wave optics corrects geometric optics, relativistic 
mechanics corrects Newtonian mechanics etc.

This objection puts limiting relations between theories centre-stage:  which I have postponed 
to the companion paper. So here it will suffice to make two general points.

\indent (1): Nagel himself replied that indeed a case
in which $T_t$'s laws are a close approximation to
what strictly follows from $T_b$ should count as reduction. He called this `approximative 
reduction' (1979, pp. 361-363, 371-373). Compare also  Hempel (1965, p. 344-346; 1966, pp. 
75-77). In a similar vein, Schaffner (1967, p. 144; 1976, p. 618) requires a strong analogy 
between $T_t$ and its corrected version, i.e. what strictly follows from $T_b$). Sadly, the 
literature seems not to notice these long-past replies to Kemeny et al. (e.g. Needham 2010, 
Section 5).

Against these replies, critics have complained that it is too programmatic, since it gives no 
general account of when an approximation or analogy is close enough. But in Nagel's and 
Schaffner's defence, I would say  that (a) we
should not expect, and (b) we do not need, any such general account.\footnote{I am not alone 
in this defence: cf. Nickles (1973, p. 189, 195) and Dizadji-Bahmani, Frigg and Hartmann 
(2010: Section 3.1). In a similar vein, the latter argue (their Section 5) that Nagelian 
reduction does not need to settle once for all whether bridge laws (cf. (3) of Section 
\ref{defextdefined}) state ``mere'' correlations, law-like correlations or 
property-identities: (which are Nagel (1961)'s three options, at pp. 354-357). I entirely 
agree, {\em pace} authors such as Kim and Causey cited at the end of Section 
\ref{defextdefined}.} This is in line with both: (i) my scepticism about a single best 
concept of reduction, or of emergence (Sections \ref{avoidcontr} and \ref{middle}); and (ii) 
my swings-and-roundabouts moral about exactly which theory to consider (end of Section 
\ref{defext}).
 What {\em matters}, both scientifically and conceptually, is that in a given case we can 
deduce that $T_t$'s proposition (whether a description of particular fact, or a general 
theorem/law) is approximately true; and that we can quantify how good the approximation is.

\indent (2): This allowance that in reduction, $T_b$ need only imply an approximation, or 
``cousin'' of $T_t$, corrected by the lights of $T_b$, fits well with a central claim of the 
companion paper. There I will not only claim that in some examples, novel and robust 
behaviour is rigorously deducible in some limit of a parameter, say $N = \infty$: showing 
emergence and reduction. I will also claim that often there is a weaker sense of `novel and 
robust behaviour' which, even with the parameter $N$ {\em finite}, can be {\em deduced}. This 
weak sense is often the idea just discussed: i.e. a matter of approximate cousins, especially 
coarse-grained versions, of the properties (especially, physical quantities and states) that 
are exactly defined, and deducible, at the $N = \infty$ limit. And though these cousins are 
approximate, they can be accurate to several significant figures!

Finally, mention of significant figures prompts the following remark. I have emphasized the 
power of deduction to justify $T_t$ as approximately true by the lights of $T_b$. But I 
should register the importance for heuristics of computer simulations (especially given the 
rise of computational physics). In particular, computer simulations of $T_b$ (or models of 
$T_b$) with finite $N$ often {\em show}, regardless of deduction, the approximate behaviour 
characteristic of $T_t$---and often the approximation is very accurate. Besides, the 
deduction/simulation distinction is not so sharp; (so nor is the justification/discovery 
distinction). For: (i) as discussed already in Section \ref{middle}, to secure a deduction we 
must often adopt an approximation scheme---whose warrant is sometimes a simulation; and (ii) 
we often have some sort of error-analysis of our  computer simulation, which justifies, at 
least partially, its accuracy.

\section{The philosophical tradition about supervenience}\label{superdef}
For the rest of this paper, I turn to the relation between emergence and supervenience. As 
mentioned at the end of Section \ref{assumejargon}, the
philosophical tradition is that supervenience is a weakening of Section \ref{defext}'s notion 
of definitional extension: namely, by allowing definitions that are infinitely long, as well 
as finitely long. Since definitional extension is the core idea of Nagelian reduction, this 
tradition prompted the proposal that emergence is ``mere supervenience'', i.e. supervenience 
using at least one infinitely long definition. Since the 1970s this proposal, together with 
the idea of multiple realizability ((5) of Section \ref{defextdefined}) has dominated 
philosophical discussion of emergence, and more generally of the relations between the 
special sciences and physics.

I deny the proposal: I will present examples of supervenience without emergence, and {\em 
vice versa}. Agreed, the proposal is nowadays not as popular as it was in the 1970s and 1980s 
(for reasons reviewed in Section \ref{superinf}). But my reasons for denying it are different 
from those in this debate's literature; and, I submit, more conclusive! Furthermore, my 
reasons expose two widespread faults of this literature. So it will be worth spelling out my 
reasons, and the faults they reveal.
In this Section, I discuss the philosophical tradition about supervenience as infinitary 
definitional extension. Then in Section \ref{No2}, I give my own reasons to deny the 
proposal, and I also expose the faults in the literature.

Section \ref{superinf} begins with the core idea of supervenience. This will lead to multiple 
realizability and supervenience as infinitary definition. This prompts the proposal that 
emergence is mere supervenience (Section \ref{411CisEmgceMere}). Section \ref{objsoglobal} 
describes how objections to Section \ref{superinf}'s definition of the core idea prompt one 
to consider global supervenience---which will lead in to Section \ref{No2}'s denial of the 
proposal.

\subsection{Supervenience, multiple realizability and infinity}\label{superinf}
Supervenience is normally introduced as follows. One envisages a set $O$ of objects, on which 
are defined two sets of properties, say $\cal B$ and $\cal T$; (again, `$\cal B$' for 
`bottom' and `$\cal T$' for `top'). One says that $\cal T$ {\em supervenes on}  $\cal B$, or 
{\em is implicitly defined by}  $\cal B$, (with respect to $O$), iff any two objects in $O$ 
that
match in all properties in $\cal B$ also match in all
properties in $\cal T$. Or equivalently, with the
contrapositive: any two objects that differ in a property
in $\cal T$ must also differ in some property or other in
$\cal B$. One also says that $\cal B$ {\em subvenes}  $\cal T$.

There are some standard vivid examples---which are persuasive, though of course (this being 
philosophy!) controvertible.\\
\indent (1): Let $O$ be the set of human actions, $\cal T$ their moral properties (e.g. `is a 
generous action'), and $\cal B$ their natural properties (e.g. `is a transfer of money to a 
poor relation'). (This example seems to have been first discussed by G.E. Moore and Hare, who 
introduced the word `supervenience' into philosophy.) \\
\indent (2): Let $O$ be the set of pictures, $\cal T$ their aesthetic properties (e.g. `is 
well-composed'), and $\cal B$ their physical properties (e.g. `has magenta in top-left 
corner'). \\
\indent (3): Let $O$ be the set of sentient animals, $\cal T$ their mental properties (e.g. 
`sees an edge in top-left visual field'), and $\cal B$ their physical properties (e.g. `has 
an an active edge-detector cell in the visual cortex').

These examples and others like them bring out the idea of {\em multiple realizability}: that 
the instances in $O$ of a property $P \in {\cal T}$ are very varied as regards their 
properties in $\cal B$. We say that $P$ is `multiply realized' with respect to $\cal B$; 
(other jargon: the instances are `heterogeneous').

The idea of multiple realizability has been developed in a broadly ``anti-reductionist'' 
direction, in two main ways. The first, called the `multiple realizability argument against 
reductionism', is informal (Section \ref{411Amra}). The second is formal, in that it appeals 
to the idea of infinity (Section \ref{411Binfinitevariety}). I will use the first as a 
spring-board to briefly discuss the philosophical issues about explanation etc. in reduction 
which I postponed in the preamble of Section \ref{No1}. But (true to my enthusiasm for 
reduction!) I will cite what I believe is a definitive refutation of the argument. Then I go 
on to emphasise the second way, the appeal to infinity. For it is this which leads to the 
proposal that emergence just is supervenience---which I take up in Section 
\ref{411CisEmgceMere}.

\subsubsection{The multiple realizability argument}\label{411Amra}
This argument returns us to topics like explanation, property-identity and laws. It goes, 
roughly, as follows. At least in some examples, the instances of $P$ are so varied that even 
if there is an extensionally correct definition of $P$ in terms of $\cal B$, it will be so 
long and-or disjunctive and-or heterogeneous that:\\
\indent \indent (a): explanations of singular propositions about instance of $P$ cannot be 
given in terms of $\cal B$, whatever might be the details about the laws and singular 
propositions involving $\cal B$; (similarly, if one conceives explanations in terms of events 
or occurrences: a theory using $\cal B$ cannot give an explanation of a singular occurrence 
of $P$); and-or; \\
\indent \indent (b): $P$ cannot be a natural kind, and-or cannot be a law-like or projectible 
property, and-or cannot enter into a law,  from the perspective of $\cal B$.\\
\indent Usually an advocate of (a) or (b) is not ``eliminativist'', but rather 
``anti-reductionist''. $P$ and the other properties in $\cal T$ satisfying (a) and-or (b) are 
not to be eliminated as a cognitively useless, albeit conveniently short, {\em facon de 
parler}. Rather, we should accept the taxonomy they represent, and thereby the legitimacy of 
explanations and laws invoking them. Probably the most influential advocates have been: 
Putnam (1975) for version (a), with the vivid example of a square peg fitting a square hole, 
but not a circular one; and Fodor (1974) for version (b), with the vivid example of $P$ = 
being money.\footnote{The argument is also influential in discussions about limiting 
relations between physical theories, especially the significance of ``singular'' limits. For 
example, I read Batterman's many passages stressing the need for explanations that abstract 
from microscopic details as rehearsing version (b) of the multiple realizability argument. 
For the (a)/(b) contrast corresponds to Batterman's distinction between explaining an 
individual instance of a pattern and explaining why the pattern can in general be expected; 
and Batterman's passages link the latter to multiple realizability, either under that name, 
or under physicists' near-synonym `universality'. (Batterman calls the corresponding types of 
why-question `type (i)' and `type (ii)': 1992; 2002, p. 23.) The passages include: (2002, p. 
71-74, 119), (2008, p. 23), (2009, pp. 6, 14) . Agreed, there is of course more to 
Batterman's overall views than rehearsing version (b) of the multiple realizability argument. 
For my purposes, his most important views are: (1) that asymptotic analysis, and singular 
limits, are essential, both to explaining universality i.e. answering type (ii) questions, 
and to emergence: cf. e.g. (2002, pp. 74-76, 134-135), (2005, pp. 236, 239, 243), (2006, pp. 
891, 894, 900, 903), (2008, pp. 19-21), (2009, pp. 23-25); and (2) that some singular limits 
are physically real (2002, p. 56; 2005, pp. 235-236; 2008, p. 19). Rueger also advocates view 
(1) (2000, p. 308; 2006, pp. 344-345). I call the views (1) and (2) `important' since in the 
companion paper (2011), I will disagree! \label{Battermann}}

I believe that Sober (1999) has definitively refuted this argument, in its various versions, 
whether based on (a) or on (b), and without needing to make contentious assumptions about 
topics like explanation, natural kind and law of nature. As he shows, it is instead the 
various versions of the argument that make contentious assumptions! I will not go into 
details: even apart from my agreeing with Sober, discussing these topics and assumptions 
about them would take up too much space. (But I recommend Shapiro (2000) and Papineau 
(2010).) Suffice it to make three points, by way of summarizing Sober's refutation. The first 
two correspond to rebutting (a) and (b); the third point is broader and arises from the 
second.

\indent As to (a): the anti-reductionist's favoured explanations in terms of $\cal T$ do {\em 
not} preclude the truth and importance of explanations in terms of $\cal B$. As to (b): a 
disjunctive definition of $P$, and other such disjunctive definitions of properties in $\cal 
T$, is no bar to a deduction of a law, governing $P$ and other such properties in $\cal T$, 
from a theory $T_b$ about the properties in $\cal B$. Nor is it a bar to this deduction being 
an explanation of the law.

The last sentence of this refutation of (b) returns us to the question raised in Section 
\ref{2.2.2.A} whether to require reduction to obey further constraints apart from deduction. 
The tradition, in particular Nagel himself, answers Yes. Nagel in effect required that the 
{\em definiens}  play a role in the reducing theory $T_b$. In particular, it cannot be a very 
heterogeneous disjunction. (Recall: the {\em definiens} is the right-hand-side of a bridge 
principle. A common ontological, rather than linguistic, jargon is: $P$'s {\em subvenient} or 
{\em basal} property.) The final sentence of the last paragraph conflicts with this view. At 
least, it conflicts if this view motivates the non-disjunctiveness requirement by saying that 
non-disjunctiveness is needed if the reducing theory $T_b$ is to explain the laws of reduced 
theory $T_t$. But I reply: so much the worse for the view. Sober puts this reply as a 
rhetorical question (1999, p. 552): `Are we really prepared to say that the truth and 
lawfulness of the higher-level generalization is {\em inexplicable}, just because the ... 
derivation is peppered with the word `or'?' I agree with him: of course not!

I should also mention another response to the {\em definiens} being a disjunction: viz. to 
treat each disjunct separately. That is: some philosophers propose that each disjunct should 
be taken to be a definition, and so to provide a deduction, valid for a limited 
domain---roughly speaking, the set of objects to which the disjunct applies. Each of these 
deductions is conceptually unified in a way that the deduction using the disjunctive 
definition is not. So following the tradition described in Section \ref{2.2.2.A}, these 
philosophers say that here we have {\em local reductions}, but no {\em global reduction}.

I need have no quarrel with this response. As I have emphasized, there is surely no single 
best sense of `reduction'. And a stronger sense along these lines, requiring 
non-disjunctiveness, will unquestionably make for reductions narrower in scope. What really 
matters, scientifically and philosophically, is to assess, in any given scientific field, 
just which such reductions hold good, and how narrow they in fact turn out to be. Cf. also 
Sober (1999, pp. 558-559).

I mention this response partly because it has been
much discussed  by metaphysicians as regards the mind-body relation, often using `pain' as 
the property $P \in {\cal T}$. Thus Kim, perhaps the most tenacious and prolific critic of 
the multiple realizability argument for non-reductive materialism, argues that if pain is 
disjunctive, `it is not the sort of property in terms of which laws can be formulated', and 
`multiple local reductions, rather than global reductions, are the rule' (1992: 16, 20). (Cf. 
also many other works e.g. Kim (1999, p. 138; 2005 pp. 24-26, 110-112).) Nagel (1965, p. 340, 
pp. 351-352), is an early statement of this response. No doubt this response seems more 
plausible, the greater the conceptual disparity between the families of properties 
(``levels'') $\cal T$ and $\cal B$. In particular, it seems plausible for pain with its ``raw 
feel'' contrasted with e.g. neurophysiological properties. Recall the tower of levels 
mentioned in (ii) of comment (5) on functional definitions in Section 
\ref{defextdefined}.\footnote{Of course, there is a lot more to Kim's overall position about 
materialism (1998, 2005) than advocating this response: Kim (2006, pp. 550-555, 557-558) is a 
recent survey, cast partly in terms of our topic, emergence. Horgan (1993, p. 575-577) is a 
fine brief survey of the prospects for non-reductive materialism. Marras (2002) and Needham 
(2009) are recent critiques of Kim.}

So much by way of discussing the multiple realizability argument. I turn to the second, more 
formal, way in which multiple realization has been used to support ``anti-reductionism''.

\subsubsection{Infinite variety}\label{411Binfinitevariety}
Some philosophers make the idea of multiple realizability more precise in terms of infinity. 
They say that, on any exact definitions of $O, {\cal T}, {\cal B}$ (in the example at 
issue---moral, aesthetic, mental etc.), there are properties $P \in {\cal T}$ with no {\em 
finite} definition in terms of $\cal B$. That is: for at least one such $P$, there are 
infinitely many $\cal B$-ways to have $P$, that are exemplified in $O$.

This meshes well with our original definition of supervenience. For one can argue that the 
definition is equivalent to an
infinitary analogue of definitional extension: in which one allows a {\em definiens\/} used 
in
giving a definitional extension to be infinitely long. The idea of allowing an infinite
sequence of compounding operations is at first sight
mind-stretching; and one worries that technical obstacles,
even paradoxes, may be lurking. But if one takes as one's
stock of operations, just the logical operations---the case
on which the logic and metaphysics literature have
concentrated---the idea is well understood. And for this equivalence, one needs ``only'' the 
idea of an infinite disjunction, as follows. Suppose that supervenience as I defined it 
holds. Then
for each property $P \in \cal T$, one can construct a
definition of $P$ (as applied to $O$) by taking the
disjunction of the complete descriptions in terms of $\cal B$ of all the objects in $O$ that 
instantiate
$P$. This disjunction will indeed be infinite if there are
infinitely many ways objects can combine properties in
$\cal B$ while possessing $P$. But in any case, supervenience
will ensure that objects matching in their complete
$\cal B$-descriptions match as regards $P$. So the
disjunctive {\em definiens\/} is indeed coextensive with
$P$.

Terminology: in the light of this equivalence,
philosophers sometimes call supervenience `infinitary
reduction'. Here `infinitary' means `infinite or finite',
and `reduction' just connotes `definitional
extension': no further  constraints (e.g. Nagel's (ii), Section \ref{2.2.2.A}), or subvenient 
properties being non-disjunctive) are
required.

This equivalence prompted the proposal that ``mere supervenience'', i.e. a reduction with at 
least one $P \in {\cal T}$ having an infinite definition in terms of $\cal B$, captures the 
relation between $\cal T$ and $\cal B$ that appears to be common across the examples (1) to 
(3): a relation where $\cal T$ is autonomous from, yet also grounded in or underpinned by, 
$\cal B$.

This proposal was widely endorsed in the 1970s and 1980s. Of course the proposal was not 
specific to these examples, and philosophers differed about which cases exemplified it. In 
particular, philosophers of science were mostly unconcerned with the moral and aesthetic 
examples (1) and (2), and focussed on (3). More generally, they endorsed the proposal as 
giving the key relation between a special science and fundamental science. That is: for many 
philosophers, mere supervenience seemed to capture the way a special science such as 
psychology (or better: a specific theory in such a science) was autonomous from, yet also 
grounded in, a basic science such as physics (or better: a specific theory in it, such as the 
physical theory of the brain). I have not yet used the word `emergence' in this connection, 
but of course many discussions did. Agreed, they often construed `emergence' differently from 
my construal using novel and robust properties or behaviour. (Often, they used causal 
notions: e.g. properties in $\cal T$ had to be causally efficacious.) Thus the proposal was 
that emergence (maybe construed differently from me) is mere supervenience. From now on, 
this proposal will be my main focus---and target.

\subsection{Is emergence mere supervenience?}\label{411CisEmgceMere}
One attractive feature of this proposal is evident from the multiple realizability argument; 
i.e. from the controversies about the role of explanation, property-identity etc. in 
reduction. Namely, the proposal promises to cut through these  controversies with the sharp 
distinction between finitude and infinity.  In short, it suggests:\\
\indent (i): If $T_t$ is a finite definitional extension of $T_b$---each {\em definiens} 
finite, be it ever so long---then there is reduction; and on the other hand:\\
\indent (ii): if at least one definition is infinite, i.e. there is mere supervenience, then 
there is emergence.

I have been anti-essentialist about the meanings of the words `reduction' and `emergence', 
and have for the most part conceived reduction in terms of definitional extension. These 
views, and the promising feature just mentioned, of course incline one to endorse the 
proposal. But I claim that it is false.

In part, my reasons against the proposal are just endorsements of points made in the 
literature  since the late 1980s; and so they apply to the proposal using construals of 
emergence different from mine. But I also have reasons of my own, which are specific to my 
construal of emergence as novel and robust behaviour, and are (say I!) conclusive. So to save 
space and avoid repetition, I will not go into detail about my endorsements of others' 
points, but will concentrate on my own reasons. As I mentioned in this Section's preamble, 
these reasons have the further significance (say I!) of revealing two faults which are 
widespread in the literature.

So the overall plan of battle for the rest of the paper is as follows.
Think of the proposal as a pair of implications, from mere supervenience to emergence and 
{\em vice versa}: I will deny both implications.\\
\indent (1):  In Section \ref{2.4.3.A}, I will deny the first implication: I will give cases 
of mere supervenience without emergence (i.e. without novel properties). These cases are 
built on a theorem  of logic (Beth's theorem) reviewed in Section \ref{Beth}, that has been 
neglected in the philosophical  literature. And this neglect is the literature's first error: 
for the theorem  collapses the proposal's crucial notion of mere supervenience! To explain 
this, we need to be more precise about how to define supervenience than my original 
definition, at the start of this Subsection, was. Section \ref{objsoglobal} will begin by 
improving the definition, and this will lead us back (in Section \ref{Beth}) to 
supervenience's being equivalent to  infinitary  definitional extension, and thereby to 
Beth's theorem and the collapse.\\
\indent (2):   Finally, on the other hand: in Section \ref{2.4.3.B}, I will deny the second 
implication. That is, I will give cases of  emergence (i.e. novel and robust behaviour) 
without mere supervenience. These cases will reveal the second fault: a tendency to wrongly 
endorse a supervenience thesis of micro-reductionism.

But first, I should close this Subsection by reporting one of the recent literature's main 
reasons (which I endorse) against the proposal that emergence is mere supervenience. For this 
reason is independent of Section \ref{objsoglobal}'s improving the definition of 
supervenience. And more important, it connects my position about emergence to a widespread 
current tendency to take supervenience as a necessary condition of emergence, and then 
propose further conditions.  (I mentioned this tendency in Section \ref{emcenr}: the proposed 
further conditions include ideas like computational intractability and paradigms like 
cellular automata.)

The  reason is essentially that because mere supervenience is a formal relation, it is 
compatible with various different conceptions of how properties in $\cal T$ are related to 
properties in $\cal B$: that they are wholly different from those in $\cal B$ though 
correlated with certain (infinitary) compounds of them, or that they {\em are} such 
compounds, or that they are to be eliminated in favour of such compounds. Indeed, this 
compatibility claim is very familiar, for both the case of morality and of mentality 
(examples (1) and (3) at the start of Section \ref{superinf}). And it is familiar for {\em 
finitary} definitional extension as much as for infinitary (i.e. as for mere supervenience). 
Thus in moral philosophy, Moore adopted the first ``realist'' conception I just listed: moral 
properties supervene on, but are wholly different from, natural properties; while Hare, on 
the other hand, was an ``irrealist''.  To see the claim in a bit more detail, consider the 
case of mentality, and a finite explicit definition of a mental predicate $P$ in terms of 
physical ones, i.e. a statement that $P$ is co-extensive with a physical predicate (no doubt 
a very complex one). And let us use the jargon of possible worlds; so we suppose the 
co-extension is non-accidental, i.e. it holds in each of a ``large'' class of possible 
worlds. These suppositions still leave open the question whether there is an identity of {\em 
properties}. This is so even if the co-extension is ``nomic'' in some sense, e.g. holding in 
all worlds that have the same laws of nature as the actual world. And some would say that 
property-identity is not settled even by necessary equivalence, i.e. by co-extension in all 
metaphysically possible worlds. (Discussions pressing this compatibility claim as showing 
that supervenience settles little include: Horgan (1993, pp. 560-566), Stalnaker (1996, pp. 
222-225), Kim (1998, p. 9; 2006, pp. 555-556) and Crane (2010, p. 26, p. 28).)

For my topic of emergence, the significance of this compatibility claim is that while 
philosophers differ about how to use `emergence', each expects it to provide some unified 
conception of the relation between the families of properties $\cal T$ and $\cal B$ that goes 
beyond metaphors like `autonomy but groundedness'. So since rival conceptions like those of 
Moore and Hare can be combined with mere supervenience, it surely cannot fit the bill. Hence 
the search, by philosophers intent on a general characterization of emergence, for other 
conditions: usually mere supervenience is seen as a necessary condition, and the search is 
for {\em further} conditions. Some proposals appeal to the ideas of the British emergentist 
Broad (van Cleve (1990, p. 222f.; McLaughlin (1997, pp. 90-96)), or are reminiscent of his 
configurational forces mentioned in footnote \ref{Broadn} (Humphreys 1997, Section 5, pp. 
117-121). Some proposals appeal to the ideas and paradigms (i) and (ii) listed in Section 
\ref{middle}. And some proposals appeal to singular limits (cf. Batterman's and Rueger's view 
(1) in footnote \ref{Battermann}).

My own response to this tendency in the literature is as I  said in Section \ref{middle}. I 
agree that emergence is not supervenience, though for reasons of my own (Section 
\ref{SEindpdt}). And I am not essentialist about how to use `emergence'; so I can be 
pluralist and accept that, for example, computational intractability in cellular automata, 
gives an acceptable sense of `emergence'. (This example is---in shortest possible 
form!---Bedau's proposal. Humphreys suggests a friendly amendment (2008, pp. 435-437, thesis 
2); a concordant scientific result, about modelling phase transitions, is  Gu, Weedbrook et 
al. (2008).) More specifically, in the companion paper's four examples of emergence and 
reduction in physics, the emergence  turns on taking a limit of some parameter. So again, 
taking such a limit may also give an acceptable sense.

\subsection{Relations and possibilia: local vs. global supervenience}\label{objsoglobal}
I now start on the plan of battle announced in Section \ref{411CisEmgceMere}. I begin with 
two related objections to Section \ref{superinf}'s opening definition of supervenience.\\
\indent (i): The definition  mentions properties, and matching in them, but not relations 
with two or more argument-places. But surely we should include {\em relations}, and somehow 
make sense of objects $o_1, o_2 \in O$ matching in them? Indeed, Section \ref{superinf}'s 
opening examples (1) to (3) support this, especially if we think of properties as intrinsic. 
The moral properties of an action surely depend on a wide array of relations it has to other 
items; similarly, the mental properties of an animal (as in theories of wide content, in the 
philosophy of mind); and similarly, the aesthetic properties of a picture. \\
\indent (ii): Supervenience's conditional, `if $o_1, o_2$ match as regards $\cal B$, they 
match as regards $\cal T$' is usually taken, for simplicity, as a mere material conditional. 
But then, for the obvious choices of the set $O$ as {\em actual} objects, the antecedent will 
very likely be false, and supervenience will be merely vacuously true. For example, it is 
very likely that no two actual pictures match in their physical properties; even allowing for 
photographs and forgeries, there will be invisible, at least microscopic, differences. \\
\indent These two objections support each other in the obvious way: including relations, as 
(i) suggests will make matching even rarer, aggravating (ii).

Both these objections can be met, at least for the most part, by a common response which we 
might call `going modal'. Suppose first that we take $O$ to contain not just actual objects 
(of a given sort: actions, pictures etc.) but also {\em possible} objects. That should answer 
(ii): for surely, there could be two actions (or pictures, or animals) that were 
atom-for-atom duplicates of each other, either both in the same possible world or in two 
different worlds. (A supervenience claim that considers trans-world, as well as intra-world, 
matching as regards $\cal B$ will be stronger than one that only considers intra-world 
matching.) But once modality is in play, it is natural to suggest that we take $O$ to contain 
{\em possible worlds}, i.e. extended, indeed global, possible states of affairs (patterns of 
fact), so that supervenience is about the matching of entire worlds as regards the 
description of their objects by both properties and relations. That should answer (i): 
considering all the various objects in a world will ensure that we allow for $P \in {\cal T}$ 
depending on a wide array of $\cal B$-relations.

This suggests that we formulate supervenience along the following lines, using a set of 
possible worlds $O$:
\begin{quote}
For all worlds $w, w' \in O$: if $w, w'$ match as regards the descriptions of their objects 
by the properties and relations in $\cal B$, then they also match as regards the descriptions 
of their objects by the properties and relations in $\cal T$.
\end{quote}
Here one envisages that worlds matching as regards the description of their objects will be 
made precise in terms of isomorphism (a bijection between the worlds' sets of inhabitants 
that preserves properties and relations); or more metaphysically, in terms of duplicate 
worlds.

Some jargon (due to Kim (1984)): taking the set $O$ to
contain possible worlds is often called {\em global supervenience}, while taking $O$ to 
contain objects (actual and-or possible) is {\em local supervenience}.  Of course, different 
authors' formulations vary somewhat; even for a single choice of $O$ as e.g. all worlds 
sharing the actual world's laws of nature. For example: (i) as regards global supervenience, 
formulations vary about whether  the same bijection must be used for the $\cal T$-isomorphism 
as for the $\cal B$-isomorphism; and (ii) as regards local supervenience, formulations vary 
about whether to consider trans-world matching of objects. Discussions covering (i) and (ii) 
include: Kim (1984); Teller (1984, pp. 140-146); Lewis (1986, pp. 14-17); Stalnaker (1993, p. 
225-230); Chalmers (1996, pp. 411-413).\footnote{Philosophers of physics will recall that 
determinism is a global supervenience thesis, viz. with $\cal B$ describing the present (or 
past and present) and $\cal T$ describing the future; so the choice in (i) yields two 
reasonable definitions of determinism. In fact, Einstein's famous hole argument of 1913 is a 
demonstration that general relativity obeys one definition but not the other. Cf. the 
definitions Dm1 and Dm2 in Butterfield (1989, pp. 7-9).}

For philosophers not sceptical of modality, global supervenience formulated along these lines 
is at least as plausible as local supervenience, i.e. object-object supervenience: (including 
the (stronger) trans-world version of local supervenience). In particular, it does not imply 
object-object
supervenience, even in the weaker intra-world version: (nor therefore can it imply its 
strengthening, the trans-world version). For global supervenience allows that two objects in 
a world, $w$ say, match each other as regards $\cal B$, but not $\cal T$, and so violate 
object-object supervenience---provided this violation is duplicated in worlds that duplicate 
$w$.\footnote{Horgan (1982, p. 40; 1993, p. 570) and Kim (1984; 1993, p. 277-278) object that 
global supervenience is too weak, since it is compatible with large and important differences 
in $\cal T$ being dependent on trivial and surely unimportant differences in $\cal B$. For 
example: surely a sensible materialism should not allow mental properties' extensions on 
planet Earth to depend on a slight displacement of a lone hydrogen atom somewhere in deep 
space. Accordingly, they go on to propose (two different) strengthenings of global 
supervenience. But I agree with Stalnaker (1996, p. 229-230) that we should rest content with 
global supervenience: `one should not define materialism so that there cannot be silly 
versions of it'; cf. also Paull and Sider (1992, pp. 841-847).}

Besides, almost all philosophers, even if sceptical of modality, accept the framework of 
classical semantics; and {\em that} framework suffices to state global supervenience very 
precisely---cf. the next Section.

\section{Emergence as mere supervenience? No!}\label{No2}
With Section \ref{superdef}'s review in hand, I can now give my own reasons for denying that 
emergence is mere supervenience. As I announced in the preamble to Section \ref{superdef}, 
this will also expose two faults in the literature. In Section \ref{Beth}, I will make 
precise Section \ref{objsoglobal}'s  notion of global supervenience. Then I point to the fact  
that a theorem in logic, Beth's theorem, collapses the distinction between supervenience thus 
defined and traditional, i.e. finitary, definitional extension---under widely endorsed 
assumptions. Neglecting this theorem is the first of the two faults I see in the literature: 
the importance of the theorem for this debate was pointed out by Hellman and Thompson already 
in 1975. Finally in Section \ref{SEindpdt}, I present examples of supervenience without 
emergence, and {\em vice versa}. This rebuts the proposal and completes my argument for the 
logical independence of emergence, supervenience and reduction. It will also expose a second 
fault of the literature: a tendency to wrongly endorse a supervenience thesis, about the 
macroscopic supervening on the microscopic, as true, when in fact it is crucially vague---and 
one salient precise version of the thesis is clearly false. Exposing this second fault will 
also return us to points [1] and [2] at the end of Section \ref{defext}.

\subsection{Supervenience often collapses into reduction}\label{Beth}
Section \ref{precise} will state Section \ref{objsoglobal}'s notion of global supervenience 
in terms of classical semantics or model theory. This will lead to the collapse to 
definitional extension, and thus to comparison with Section \ref{No1}'s topic of reduction; 
(Section \ref{precise2}).\footnote{The following discussion is based on Hellman and Thompson 
(1975, Section II), though adding a few details.}

\subsubsection{Global supervenience made precise}\label{precise}
Thus recall that a formal language $L$ has a non-logical vocabulary, which for simplicity we 
can take to comprise only predicates. That is: we suppose we have eliminated names and 
function-symbols in terms of predicates using Russell's theory of descriptions (cf. (3) of 
Section \ref{defextdefined}). (This tactic is merely for simplicity: by construing 
`definition' in a more cumbersome way, i.e. by disjunctively treating predicates, names and 
function-symbols as three separate cases, we could avoid it; cf. e.g. Boolos and Jeffrey 
(1980, p. 246).) We envisage that $L$'s set of predicates is the union ${\cal B} \cup {\cal 
T}$ of the disjoint sets $\cal B$ and $\cal T$. Among the interpretations, $I, J$ etc., of 
the language, we define in the usual way two relations: isomorphism, written $I \cong J$; 
and restriction to a sub-vocabulary, written $I|_{\cal B}, J|_{\cal T}$ etc. (Similarly, we 
will write $L|_{\cal B}, L|_{\cal T}$.)

 Now suppose we are given a class $\alpha$ of interpretations of $L$. Then there are two 
salient relations between $\cal B$ and $\cal T$, relative to the class $\alpha$.

\indent (Exp): {\em Explicit definability}: For each predicate $P \in {\cal T}$, of 
polyadicity $n$, there is an open sentence $\phi(x_1,...,x_n)$ in $L$, with only predicates 
in $\cal B$ and with $n$ free variables, such that every interpretation $I \in \alpha$ makes 
true the universally quantified biconditional, i.e. the statement of coextension: $(\forall 
x_1)...(\forall x_n)(P(x_1,...,x_n) \equiv \phi(x_1,...,x_n))$.

\indent (Imp): {\em Implicit definability}: For all $I, J \in \alpha$, if $I|_{\cal B} \cong 
J|_{\cal B}$, then $I \cong J$. (Of course, the consequent concerns $\cal T$ as well as $\cal 
B$.)

The threatened collapse of supervenience into reduction will turn on the fact  that (Exp) and 
(Imp) are provably equivalent---under widely endorsed assumptions. This is {\em Beth's 
theorem}. But let us first comment on their respective relations to our previous discussions.

(Exp) returns us to Section \ref{No1}'s notion of reduction. For the set of (Exp)'s 
biconditionals is of course the set $D$ of judiciously chosen definitions in terms of which 
(3) of Section \ref{defext} defined $T_t$ being a definitional extension of $T_b$. To spell 
out the connection more precisely, we need the idea of extensionality, as follows.\\
\indent Recall that the {\em principle of extensionality}, applied to an $n$-place predicate 
$P$, says the following. Suppose that in an interpretation $I$, $P$ is coextensive with an 
$n$-place open sentence $\phi$, i.e. $I$ makes true the universally quantified biconditional 
$(\forall x_1)...(\forall x_n)(P(x_1,...,x_n) \equiv \phi(x_1,...,x_n))$. Let $\Psi(P)$ be 
any formula containing $P$ and let $\Psi(\phi)$ be the formula obtained from $\Psi(P)$ by 
everywhere substituting $\phi$ for $P$ (with the corresponding argument-variables). Then $I$ 
also makes true $\Psi(P) \equiv \Psi(\phi)$. In short, as a rule of inference: From $(\forall 
x_1)...(\forall x_n)(P(x_1,...,x_n) \equiv \phi(x_1,...,x_n))$, infer $\Psi(P) \equiv 
\Psi(\phi)$.\\
\indent So suppose now that $L$ is an extensional language; (as Hellman and Thompson, and 
Beth's theorem, explicitly assume; as do many philosophical discussions of reduction, though 
often implicitly by referring to `formalized languages', `first-order languages' without 
further details). Let $D$ be a set of biconditionals, defining each $P \in {\cal T}$ in terms 
of $\cal B$, as in (Exp). And for any theory (indeed: mere set) $T$ of formulas of $L$, let 
$T^D$ be obtained from $T$ by everywhere substituting  for each  $P \in {\cal T}$ its 
definition $\phi$, as given by the set $D$ of definitions. So formulas in $T^D$ contain no 
predicates $P \in {\cal T}$; they are in the language $L|_{\cal B}$. $T^D$ is sometimes 
called a {\em definitional equivalent} of $T$; and similarly for formulas in $T$. Then, by 
extensionality: among the (interpretations of $L$ that are) models of $D$, any interpretation 
making true $T$ also makes true $T^D$; and vice versa. In the usual notation, writing $\Psi$ 
for any formula of $T$ and $\Psi^D$ for its definitional equivalent: $D \models \Psi \equiv 
\Psi^D$. In particular, this applies to $T := T_t$. So for any $\Psi \in T_t$: $D \models  
\Psi \equiv \Psi^D$.

Now we can return to our original concern: the assertion that $T_t$ is a definitional 
extension of $T_b$. In Section \ref{defext} this was defined in terms of derivability, i.e. 
syntactic consequence. But assuming the underlying logic is complete, we can instead use 
semantic consequence, so that the assertion is that $D, T_b \models  T_t$; or equivalently, 
that $D, T_b \models  T_t^D$. Of course, logic alone cannot guarantee that for a given choice 
of $\cal T$ (and so $\cal B$ as the rest of $L$'s non-logical vocabulary), and of $T_t$ and 
$T_b$, there is a set $D$ of definitions of all $P \in {\cal T}$, such that  $D, T_b \models  
T_t$.\footnote{
And of course some of the definitions might be so long or complicated as to violate some 
philosophers' proposed informal constraints on reduction; cf. Section \ref{2.2.2.A} and the 
multiple realizability argument in Section \ref{411Amra}.}

Turning to (Imp), it clearly expresses the idea of global supervenience. But I should notice 
two wrinkles in relation to Hellman and Thompson's discussion. First, their definition (1975, 
p.559) uses identity, not isomorphism, of models: `=' not `$\cong$'; (as do other treatments, 
e.g. Boolos and Jeffrey (1980, p. 246)). My (Imp) uses isomorphism  in order to fit better 
with our informal motivation, that global supervenience is about two possible worlds' 
matching as regards a set of properties or predicates. But as they remark (their footnote 
14), the two definitions are equivalent if $\alpha$ is closed under automorphic images: which 
it will be in the cases relevant to the collapse of supervenience to reduction, i.e. in the 
cases covered by Beth's theorem.

 Second, Hellman and Thompson also discuss a second version of global supervenience, in 
effect concerning truth rather reference. So they call (Imp) `reference-determination' and 
their other version `truth-determination'; (viz. their (4), p. 558, using the relation 
between interpretations of elementary equivalence, i.e.  two interpretations making the same 
sentences true). But I can ignore this second version; partly because, for the cases that 
mostly concern us (including the collapse of supervenience into reduction), $\alpha$ is the 
class of models of a theory---and for such $\alpha$, (Imp) implies this second version (as 
Hellman and Thompson note, p. 560).

\subsubsection{The collapse: Beth's theorem}\label{precise2}
I turn to the implications between (Exp) and (Imp), for the case where the set of 
interpretations $\alpha$ is a theory-class, i.e. the set of models of some theory $T$, and 
the language $L$ is first-order, finitary (i.e. has finitely many predicates, and finitely 
long formulas) and extensional. Then (Exp) implies (Imp). Agreed, since the definitions $D$ 
in (Exp) evidently fix the extensions of $P \in {\cal T}$, this implication is unsurprising; 
and it is straightforward to show (e.g. Boolos and Jeffrey 1980, p. 247-248).  Of course, 
this implication fosters the philosophical tradition that supervenience is a weakening of 
reduction; (as did Section \ref{411Binfinitevariety}'s equivalence of local supervenience 
with possibly-infinite disjunction).

But the converse is also true: (Imp) implies (Exp). This {\em is} surprising, and is hard to 
show. It is Beth's theorem (1953). It depends on the compactness theorem and Craig's 
interpolation lemma; (cf. Boolos and Jeffrey (1980, p. 248-249), Mostowski (1966, p. 127)). 
This dependence also means there is no limit to how long or complicated the explicit 
definitions are. Here, of course, lies a way of reconciling Beth's theorem with Section 
\ref{superinf}'s opening intuition that aesthetic properties of pictures, or mental 
properties of animals, supervene on, but are not definable from, their physical properties. 
Namely:  if a definition is allowed to be a billion pages long (or much longer!), nobody can 
be so sure that there is none!

In the light of this equivalence of (Exp) and (Imp), there are two obvious broad tactics 
available to someone seeking to define some relation of dependence, between a theory that is 
`higher-level' or part of a `special
science', and a `lower-level' or `fundamental' theory, that is weaker than reduction.

 First, one can accept the equivalence, but argue, as in Section \ref{2.2.2.A} and the 
multiple realizability argument of Section \ref{411Amra}, that reduction is {\em stronger} 
than deducibility, i.e. stronger than (Exp) above. Thus (Exp), equivalently (Imp), define the 
weaker relation that is sought: deducibility indeed, though perhaps with definitions a 
billion pages long. Second, one can deny that the assumptions of the equivalence---that 
$\alpha$ is a theory-class, and $L$ first-order, finitary and extensional---apply to the 
philosophical or logical description of scientific theories and their relations: so that one 
can still advocate (Imp) as the sought-for relation.

The second tactic is the one that Hellman and Thompson (1975, pp. 562-563) adopt. They recall 
that in order to articulate global supervenience (of one theory's, or science's, taxonomy or 
set of predicates, on another's), the set $\alpha$ must represent some sort of scientific 
possibility (cf. Section \ref{objsoglobal}'s motivations for `going modal'); and they point 
out that there is good reason to {\em deny} that this $\alpha$ is a theory-class. For since 
G\"{o}del's 1931 incompleteness theorem, modern logic has taught us that most formalized 
theories (in particular, first-order finitary theories rich enough to contain arithmetic) 
have non-standard models. Since it is reasonable to exclude such models, i.e. to insist on 
certain predicates receiving their standard interpretations (either up to isomorphism of 
models, or yet more finely), it is reasonable to {\em deny} that the set of scientifically 
possible structures is a theory-class (for theories with such non-standard models).

It is clear that for my purposes, I do not need to decide between these two tactics. As to 
the first tactic, recall Section \ref{defext}'s celebration of the power of deduction, and 
its anti-essentialism about reduction, i.e. indifference as to how to use the word 
`reduction', and its alleged contraries like `emergence'. So I can concur with someone who 
reserves `reduction' for short and simple deductions, using short and unified definitions, 
and who suggests the weaker concept of deduction {\em tout court}---maybe long and 
complicated---as the more general dependence between higher-level and lower-level theories.

 As to the second tactic, I of course endorse Hellman and Thompson's point that for 
G\"{o}delian reasons, the set of scientifically possible structures is not a theory-class for 
a first-order, finitary extensional theory. This leaves the way open for them to advocate 
(Imp) as the desired weaker-than-reduction relation---which they proceed to do in another 
paper (1977).  I will not assess their position in full. (In any case, a recent discussion by 
Hellman (2010, Section 4) suggests that like me, he could also endorse the first tactic.) But 
in the next Section, I will argue for my own main claim, announced in Section 
\ref{411CisEmgceMere}: that emergence in my sense, and supervenience---now meaning global 
supervenience, i.e. (Imp)---are independent.\footnote{For higher-order or infinitary 
languages, Beth's theorem fails: which suggests that supervenience theses such as physicalism 
might best be made precise as implicit definability, {\em a la} (Imp), in a theory in such a 
language. Hellman and Thompson (1977, p. 337) give this a detailed but sceptical 
assessment---with which I concur.}

\subsection{Supervenience without emergence, and {\em vice versa}}\label{SEindpdt}
I turn to my four cases of logical independence: $E \& S$, $E \& \neg S$, $ \neg E \& S$ and 
$ \neg E \& \neg S$. As discussed already in (ii) of Section \ref{8fail}, the first and 
fourth cases are already established. That is: assuming as usual that reduction implies 
supervenience, any case of $E \& R$ (secured by Section \ref{No1} and the companion paper) is 
a case of $E \& S$. And two theories about two unrelated topics will give cases of  $ \neg E 
\& \neg S$. So I only need to consider the second and third cases: $E \& \neg S$ and $ \neg E 
\& S$.

But as explained in Section \ref{411CisEmgceMere}, I will argue for these cases in the more 
interesting sense that I will deny the proposal  that {\em all} and {\em only} cases of mere 
supervenience, i.e. supervenience-but-not-reduction,  are cases of emergence.  The `all' 
claim is false: $S \& \neg R$ does not imply $E$, so that $(S \& \neg R) \& \neg E$ is 
possible. Section \ref{2.4.3.A} will give three cases, thanks to Hellman and Thompson. Their 
third case reveals a fault in the literature; which will be relevant in Section 
\ref{2.4.3.B}'s demonstration that the `only' claim is false: $E$ does not imply $S \& \neg 
R$, so that $E \& \neg(S \& \neg R)$, i.e. $E \& (\neg S \vee R)$, is possible.

\subsubsection{Supervenience without emergence or reduction}\label{2.4.3.A}
Hellman and Thompson (1977, Section 2) give three examples of supervenience without 
reduction. I shall argue that all three involve no emergence in my sense. For all I know, my 
claim involves no disagreement with Hellman and Thompson. Though they sometimes mention 
emergence, it is not their focus---they aim rather to articulate physicalism as supervenience 
(which they call `determination') without reduction. In particular, they do not construe 
emergence as novel and robust behaviour: a construal on which my claim will turn.

Their first two examples are adapted from theorems about formal systems of arithmetic.  But 
since almost any formalized physical theory is likely to contain a version of arithmetic, the 
examples could no doubt be adapted to apply to physics. In any case, the first example is 
based on a renowned theorem (Tarski's indefinability theorem), and it leads in to the second 
example; so I will concentrate on the first. The third example is just a sketch, about 
statistical mechanics. But it will teach us a moral that will also apply in Section 
\ref{2.4.3.B}.

(1): {\em Tarski's indefinability theorem}: This theorem says that the set of G\"{o}del 
numbers of sentences true in the standard model of arithmetic is not definable in arithmetic; 
(for definitions and a proof, cf. e.g. Boolos and Jeffrey 1980, p. 173-177). So consider a 
(first-order) theory $T$ of arithmetic in a language $L$; so $L$ contains symbols for zero, 
successor, addition and multiplication. Extend $T$ by adding: (i) a one-place predicate $Tr$; 
and (ii) for each sentence $S$ of $L$ that is true in the standard model of arithmetic, with 
G\"{o}del number $n(S)$, an axiom `$Tr(n(S)) \equiv S$'. Call the extended theory $T^*$. Then 
Tarski's theorem says that $Tr$ is not explicitly definable in $T$. But on the other hand: 
taking $\alpha$ as the class of standard $\omega$-models of $T^*$, $Tr$ is implicitly 
definable in terms of $L$'s vocabulary (symbols for zero, successor etc.) relative to 
$\alpha$. For evidently: once the reference of $L$'s arithmetic vocabulary is fixed, so is 
the reference of `true-in-arithmetic'.\footnote{On the other hand, taking $\alpha$ as the 
class of {\em all} models of $T^*$, so that Beth's theorem applies: we infer from Tarski's 
theorem that $Tr$ is not implicitly definable in terms of $L$'s vocabulary, relative to this 
$\alpha$.}

\indent So much by way of summarizing Hellman and Thompson's (1977, p. 312) first example. 
Now I simply add that (by my lights at least!) $Tr$ does {\em not} represent a novel 
property, across the class $\alpha$ of standard $\omega$-models of $T^*$, in comparison with 
the properties thus expressed using $L$'s vocabulary (using the symbols for zero, successor 
etc.). Indeed, this denial goes hand in hand with the implicit definability of $Tr$, i.e. the 
fact that once the reference of $L$'s vocabulary is fixed, so is the reference of 
`true-in-arithmetic'. (Of course, I intend this hand-in-hand claim for this example and its 
ilk, like (2) below: but not as a general claim. I do not think implicit definability 
excludes novelty---as my later examples of $E \& R$, and so $E \& S$, are meant to attest.)

(2): {\em Addison's theorem}: The second example is similar to the first. Hellman and 
Thompson invoke another indefinability theorem, viz. Addison's theorem, to argue that in a 
natural class of models (though not a theory-class!), a certain predicate is not explicitly 
definable in terms of certain others, but is implicitly definable. \\
\indent To state Addison's theorem, we need the notion of a class $C$ of sets of numbers 
being definable in arithmetic. This means that in the language of arithmetic augmented with a 
monadic predicate $G$, there is a formula that is true in the standard model of arithmetic 
iff $G$ is assigned as its extension one of the sets in the class $C$. Then Addison's theorem 
says that the class of sets definable in arithmetic is not definable in arithmetic. That is: 
there is no formula in $L + G$ that is true in the standard model of arithmetic iff $G$ is 
assigned as its extension a set of numbers definable in arithmetic; (cf. Boolos and Jeffrey 
1980, p. 208 and Chapter 20, p. 217). In short: the predicate `is a set of numbers definable 
in arithmetic' is not itself definable in arithmetic.\\
\indent Hellman and Thompson then define a natural class of models, that are intuitively the 
standard models of arithmetic each augmented with the standard interpretation of `is a set of 
numbers definable in arithmetic'; (their $\alpha^C$; 1977, p. 313). Thus the predicate `is a 
set of numbers definable in arithmetic' {\em is} implicitly defined by $L$'s vocabulary (the 
symbols for zero, successor etc.) relative to $\alpha^C$.\\
\indent As in (1), I of course concur. But I would then add that (by my lights at least!) the 
predicate  does {\em not} represent a novel property, across the class $\alpha^C$, in 
comparison with the properties thus expressed using $L$'s vocabulary. Again, this denial goes 
hand in hand with the implicit definability of the predicate.

(3): {\em Statistical mechanics}: Hellman and Thompson's last example (1977, p. 314) is much 
less formal. It returns us to my example in Section \ref{defextexamples}. There my first 
point was that equilibrium classical statistical mechanics is a definitional extension of the 
classical mechanics of the microscopic constituents, taken as including Lebesgue measure on 
phase space and multi-variable calculus. But I stressed that this claim of deducibility is of 
course sensitive to the exact definitions of the theories concerned: in particular, either 
statistical mechanics should not include the ergodic hypothesis, or the microscopic theory 
should contain special assumptions to make the hypothesis deducible.\\
\indent Hellman and Thompson adapt the same circle of ideas to urge an example of implicit, 
but not explicit, definability: of supervenience without reduction (definitional extension). 
They envisage that the microscopic mechanics lacks measure theory, so that the definitional 
extension is blocked; but then say
\begin{quote}
the macro-concepts of statistical mechanics are {\em determined} by the microscopic ones: fix 
two closed systems of particles identically in micro-respects, and their macro-behaviours 
will be indistinguishable. Each system will be represented by the same trajectory in phase 
space. If, therefore, higher entropy regions, say, are entered at given times by one system, 
the same regions will be entered at those times by the other ... a clear case of 
determination. (ibid.)
\end{quote}
Broadly speaking, I concur: at the cost of considerable labour, one could formalize both the 
microscopic and the macroscopic concepts and theories, so as to give a rigorous example of 
implicit, but not explicit, definability. But what about emergence? Admittedly, Section 
\ref{defext}'s eulogy for reduction as deduction urged that there are cases of emergence and 
reduction: $E \& R$. But that by no means implies that every macroscopic concept, in any 
rigorous example along the lines indicated, would be emergent. Surely not! Examples (1) and 
(2) have shown that even elementary  arithmetic harbours implicitly, but not explicitly, 
definable concepts that are not novel. So we can surely find such concepts in the much richer 
setting of  rigorous statistical mechanical examples.

\indent This quotation from Hellman and Thompson reveals the second fault I promised. It 
turns on the fact that they refer to fixing a system's `micro-respects', but they do not 
specify what these properties are. Agreed, that is fair enough when sketching an example. But 
we should note that countless philosophical discussions of supervenience, emergence and 
related issues, similarly fail to specify the `micro-respects' they are concerned with, not 
just when sketching an example, but in their official statement of doctrine. For example, 
countless discussions of the supervenience of the mental on the physical (example (3) of 
Section \ref{superinf}) refer to an animal in some physical state, and an `atom-for-atom 
replica/duplicate' of the animal: with no precise statement of which properties must match 
between such replicas. This lacuna might seem a minor matter; or even a judicious openness to 
whichever properties physics might discover (whether yesterday, today or tomorrow). But it 
obviously threatens to render supervenience claims trivially true, by the set $\cal B$ of 
subvening properties or predicates being taken as {\em however} large and inclusive as is 
needed to fix the properties or predicates in $\cal T$. As we will see in Section 
\ref{2.4.3.B}, this fault tends to hide cases that are {\em failures} of supervenience, 
including those that are also cases of emergence.\footnote{Of course I am not the first to 
notice the danger of supervenience claims becoming trivial owing to vague inclusiveness about 
the subvening basis $\cal B$; and the similar danger for reducibility claims. Healey (1978) 
and Crane and Mellor (1990, Section 2) discuss the danger that physicalism is trivially true 
owing to the elasticity of the term `physics'. I myself think `physics' has a substantial 
enough meaning to avoid the danger (1998, Section 3.1.3, p. 133).}

Finally, it is worth noting that, apart from this philosophical danger of trivializing 
supervenience, there is a corresponding scientific danger. Namely, complacency about 
supervenience (or reduction): saying that somehow---we need not care how!---the microscopic 
properties are defined so as to subvene (or reduce) whatever phenomena we see, however 
surprising: e.g. superfluidity as in the example at the end of Section \ref{defext}.

\subsubsection{Emergence without supervenience or reduction}\label{2.4.3.B}
To complete my argument for the logical independence of emergence, reduction and 
supervenience, I turn to giving  cases of emergence without supervenience or reduction. There 
will be two sorts of case, [1] and [2], corresponding to the points [1] and [2] at the end of 
Section \ref{defext}; (so we have already done most of the work!). Given my over-arching 
message, about the compatibility of reduction and emergence, it will be no surprise to hear 
that one sort of case (the second) is {\em counterfactual}. So for these cases, I am leaning 
on my argument being for {\em logical} independence, not some sort of nomic independence (cf. 
Section \ref{assumejargon}). But anyway, the first sort of case is both real and pervasive.

[1]: At the end of Section \ref{defext}, I noted the uniformity of the rules, in classical 
and quantum physics, for defining a composite system's state-space and its quantities; viz. 
for state-spaces, Cartesian products and tensor products respectively. The uniformity of the 
rules served my purpose there: namely eulogizing the power of reduction.\\
\indent But on the other hand, several philosophers have argued that the quantum rules 
harbour a very different moral: namely, that the existence of {\em entangled states} in the 
tensor product of two Hilbert spaces, makes for important, indeed pervasive, cases of 
emergence combined with a failure of supervenience (and so of reduction).\footnote{I will not 
quote the definitions of supervenience, usually called `mereological supervenience'. For 
details cf. Healey (1991, especially Section V, p. 408 onwards), Humphreys (1997, p. 122), 
Silberstein (2001, pp. 73-78; 2002, pp. 96-98), Silberstein \& McGeever (1999, p. 187-189), 
Howard (2007).} What am I to make of this?

I can simply agree with these authors. Of course, we have no disputes about quantum physics' 
formalism in itself. Nor should (or could!) we disagree about the sheer logic of how a 
precisely described case gets classified as emergent, supervenient or whatever. So, like in 
Section \ref{defextexamples}'s example of reducing statistical mechanics to micro-mechanics,  
it is a case of swings and roundabouts: we just have to decide what precise theories (and-or 
vocabularies, and-or classes of models), and what precise relations of emergence, 
supervenience etc. we are concerned with. Besides, I concur that entangled states violate 
their formulations of supervenience, and yield emergence i.e. novel and robust 
behaviour.\footnote{No philosopher of physics will deny that violations of Bell inequalities, 
and other phenomena teased out of entangled states by the burgeoning field of quantum 
information, are novel and robust behaviour. Agreed, I also need to gloss entangled states as violating supervenience formulated in my preferred way, as a relation between theories, rather than as in the previous footnote. I will not go into details; but the idea is to define a theory $T_b$ that attributes only product states, and perhaps their mixtures, to composite systems. Then much of the usual theory, e.g. its attribution of correlations between component systems, will not supervene on $T_b$. Thanks to a referee for raising this point.} Accordingly, these states give cases of 
emergence without supervenience.

 [2]: At the end of Section \ref{defext}, I noted that according to both classical and 
quantum physics, there are no fundamental forces that come into play only when the number of 
bodies (or particles, or degrees of freedom) exceeds some number, or when the bodies etc. are 
in certain states. Indeed, only two-body forces are needed. And I mentioned that McLaughlin 
(1992) claims that quantum physics'  not needing such forces, even for its explanations of 
chemical and biological phenomena, spelt the end of British emergentism of Broad's stripe. 
(Just before the rise of quantum chemistry, Broad had conjectured that such forces, which he 
called `configurational forces', would be needed to explain such phenomena.)

\indent But (here I ``go counterfactual''): suppose there were such configurational forces, 
and that they were needed to  explain some chemical or biological phenomenon or phenomena. 
This supposition is perfectly reasonable: recall from the end of Section \ref{defext} how 
Leggett in the 1970s suspected such forces were needed to explain superfluid helium, but then 
discovered they were not. Reasonable, though {\em false}: {\em pace} some (surely maverick?) 
philosophers of chemistry who claim that phenomena such as molecular shape require such 
forces (e.g. Hendry 2010, Section 3).

On this supposition, then: no matter what the exact details were, we would surely have a 
second class of cases of emergence without supervenience. For the chemical or biological 
phenomenon would give the novel and robust behaviour that is emergence. But the need for 
configurational forces would mean that the phenomenon does {\em not} supervene on all the 
quantum-physical properties and facts that involve only two-body forces; (i.e. the actual 
properties and facts!).

\indent Here we return to the fault revealed at the end of Section \ref{2.4.3.A}; and the 
moral it teaches, that  a supervenience claim needs to define precisely what properties or 
predicates are in the subvening set $\cal B$, on pain of being trivialized by phrases like 
`atom-for-atom duplicate' being taken to cover however large and inclusive a set of 
microscopic properties (Hellman and Thompson's `micro-respects') as is needed. Thus in the 
envisaged situation, physics as a discipline would no doubt address itself to developing a 
quantitative theory of the configurational forces: a theory that would then use the forces to 
explain the chemical or biological phenomenon. And if this endeavour succeeded, we might well 
sum up the situation by saying that the phenomenon supervened on, or even was reduced to, the 
physical. That would be testimony, yet again, to the power of reduction---and perhaps to the 
elasticity of the word `physics'! But that in no way supports the claim that the phenomenon 
supervenes on the (actual) physics that uses only non-configurational forces---{\em ex 
hypothesi} that claim is false.\footnote{Of course, this moral about the need for precision 
applies  to a claim of emergence, as much as to one of supervenience: as discussed in Section 
\ref{emcenr}, there is vagueness about what counts as `novel' and `robust'. But one naturally 
takes it in one's stride that in the envisaged situation, the chemical or biological 
phenomenon counts as behaviour novel and robust enough to be emergence; not least because of 
the long and distinguished pedigree of emergentism about chemistry and biology. On the other 
hand, the triumphant successes of reduction in terms of microscopic 
constituents---especially, the achievements of statistical mechanics, and then quantum 
chemistry and molecular biology---make it all too tempting to slide from `atom-for-atom 
duplicate' being interpreted properly, in terms of some specific micro-physical theory, to it 
being interpreted trivially, i.e. as something like `whatever it takes to subvene all the 
facts'.} \\

{\em Acknowledgements}:--\\
I am indebted to many audiences and colleagues, and two referees. For invitations to lecture, I am very 
grateful to: the Royal Society, Bristol University, the Italian Society for Philosophy of 
Science, Lee Gohlicke and the Seven Pines Symposium, the Lorentz Centre at the University of 
Leiden, the University of Pittsburgh and Sandra Mitchell, and Princeton University. I am also 
grateful: to my hosts and audiences at these occasions, and at seminars in Cambridge, 
Chicago, Maryland, Minnesota, Oxford and Sydney; to Geoff Hellman, Don Howard, Paul Humphreys, Michael 
Silberstein, Elliott Sober, and Ken Schaffner for comments and-or encouragements; to Paul 
Mainwood, for discussions and the inspiration of his (2006); and to the editors, not least 
for their patience.

\section{References}

Batterman, R. (1992), `Explanatory instability', {\em Nous} {\bf 26}, pp. 325-348.

Batterman, R. (2002), {\em The Devil in the Details}, Oxford University Press.

Batterman, R. (2005), `Critical phenomena and breaking drops: infinite idealizations in 
physics', {\em Studies in History and Philosophy of Modern Physics} {\bf 36B}, pp. 225-244.

Batterman, R. (2006), `Hydrodynamic vs. molecular dynamics: intertheory relations in 
condensed matters physics', {\em Philosophy of Science} {\bf 73}, pp. 888-904.

Batterman, R. (2008), `On the explanatory role of mathematics in empirical science', 
available at: http://philsci-archive.pitt.edu/archive/00004115/

Batterman, R. (2009), `Emergence, singularities, and symmetry breaking', available at: 
http://philsci-archive.pitt.edu/archive/00004934/

Bedau, M. (2003), `Downward causation and autonomy in weak emergence', {\em Principia Revista 
Internacional de Epistemologica} {\bf 6}, pp. 5-50; reprinted in Bedau and Humphreys (2008); 
page reference to the reprint.

Bedau, M. and Humphreys, P. (eds.) (2008), {\em Emergence: contemporary readings in 
philosophy and science}, MIT Press: Bradford Books.

Bishop, R. (2008), `Downward causation in fluid convection', {\em Synthese} {\bf 160}, pp. 
229-248.

Boolos, G. and Jeffrey, R. (1980), {\em Computability and Logic}, Cambridge University Press.

Butterfield, J. (1989), `The Hole Truth' {\em British Journal for the Philosophy of Science} 
{\bf 40}, pp. 1-28.

Butterfield, J. (1998), `Quantum curiosities of psychophysics', in {\em Consciousness and 
Human Identity}, ed. J. Cornwell, Oxford University Press, pp. 122-157. PITT-PHIL-SCI00000193

Butterfield, J. (2011), `Less is Different: Emergence and Reduction Reconciled', in {\em 
Foundations of Physics}, this issue: online at Springerlink (DOI 10.1007/s10701-010-9516-1); 
and at http://philsci-archive.pitt.edu/8355/

Butterfield, J. and Isham, C. (1999), 'On the Emergence of Time in Quantum Gravity', in {\em 
The Arguments of Time}, ed. J. Butterfield, British Academy and O.U.P., 1999, pp. 111-168; 
and at: gr-qc/9901024.

Butterfield, J. (2007), 'Reconsidering Relativistic Causality', {\em International Studies in  
the Philosophy of
  Science} {\bf 21} pp. 295-328: available at: http://uk.arxiv.org/abs/0708.2189; and at:
 http://philsci-archive.pitt.edu/archive/00003469/

Causey, R. (1972), `Attribute-identities and micro-reductions', {\em Journal of Philosophy} 
{\bf 67}, pp. 407-422.

Chalmers, D. (1996), `Supervenience', in his {The Conscious Mind}, Oxford University Press, 
pp. 32-42; reprinted in Bedau and Humphreys (2008); page reference to the reprint.

Crane, T. (2010), `Cosmic hermeneutics vs. emergence: the challenge of the explanatory gap', 
in {\em Emergence in Mind}
eds. C. Macdonald and G. Macdonald, Oxford University Press, pp. 22-34.

Crane T. and Mellor D., 1990. 'There is no Question of Physicalism', {\em Mind}  {\bf 99}, pp. 
185-206; reprinted in Mellor's {\em Matters of Metaphysics} (1991), Cambridge  University 
Press.

Dizadji-Bahmani, F., Frigg R. and Hartmann S. (2010), `Who's afraid of Nagelian reduction?', 
forthcoming in {\em Erkenntnis}; available at: 
http://philsci-archive.pitt.edu/archive/00005323/

Endicott, R. (1998), `Collapse of the New Wave', {\em Journal of Philosophy} {\bf 95}, pp. 
53-72.

Feyerabend, P. (1962), `Explanation, reduction and empiricism', in H. Feigl and G. Maxwell 
(eds.) {\em Minnesota Studies in Philosophy of Science} {\bf 3}, University of Minnesota 
Press, pp. 28-97.

Field, H. (1978), `Mental representation', {\em Erkenntnis} {\bf 13}, pp. 9-61; reprinted in 
N. Block (ed.) {\em Readings in the Philosophy of Psychology} volume 1 (1980), Harvard 
University Press.

Fodor, J. (1974), `Special Sciences (Or: the disunity of science as a working hypothesis), 
{\em Synthese} {\bf 28}, pp. 97-115; reprinted in Bedau and Humphreys (2008).

Frigg, R. (2003), `Self-organized criticality---what it is and what it isn't', {\em Studies 
in History and Philosophy of Science} {\bf 34}, pp. 613-632.

Gu, M., Weedbrook C., Perales A. and Nielsen M. (2008) `More really is different', 
arXiv:0809.015 [cond-mat.other]

Healey, R. (1978), `Physicalist Imperialism', {\em Proceedings of the Aristotelian Society} 
{\bf 74}, pp. 191-211.

Healey R. (1991), `Holism and nonseparability', {\em Journal of Philosophy} {\bf 88}, pp. 
393-421.

Hellman, G. (2010), `Reduction , determination, explanation', at: www.tc.umn.edu/~hellm001

Hellman, G. and Thompson, F. (1975), `Physicalism: ontology, determination and reduction', 
{\em Journal of Philosophy} {\bf 72}, pp. 551-564.

Hellman, G. and Thompson, F. (1977), `Physicalist materialism', {\em Nous} {\bf 11}, pp. 
309-345.

Hempel, C. (1965), {\em Aspects of Scientific Explanation}, New York; the Free Press.

Hempel, C. (1966), {\em Philosophy of Natural Science}, Prentice-Hall.

Hendry, R. (2010), `Ontological reduction and molecular structure', {\em Studies in History 
and Philosophy of Modern Physics} {\bf 41}, pp. 183-191.

Horgan, T. (1982) `Supervenience and Microphysics', {\em Pacific Philosophical Quarterly}
 {\bf 63}, pp. 29-43.

Horgan, T. (1993) `From supervenience to superdupervenience: meeting the demands of a 
material world', {\em Mind} {\bf 102}, pp. 555-586.
	
Howard, D. (2007),  "Reduction and Emergence in the Physical Sciences: Some Lessons from the 
Particle Physics-Condensed Matter Physics Debate" in Evolution, Organisms and Persons.  
Nancey Murphy and Willian R. Stoeger, S.J. (eds.) Oxford: Oxford University Press.

Humphreys, P. (1997), `How Properties Emerge', {\em Philosophy of Science} {\bf 64}, pp. 
1-17; reprinted in Bedau and Humphreys (2008); page reference to the reprint.

Kemeny, J and Oppenheim P. (1956), `On reduction', {\em Philosophical Studies} {\bf 7}, pp. 
6-19.

Kim, J. (1984) `Concepts of Supervenience', {\em Philosophy and Phenomenological Research},
{\bf 45}, pp. 153-176; reprinted in Kim's {\em Supervenience and Mind} (1993), Cambridge
    University Press.

Kim, J. (1992), `Multiple realization and the metaphysics of reduction', {\em Philosophy and 
Phenomenological Research},
{\bf 70}, pp. 1-26.

Kim, J. (1998), {\em Mind in a Physical World}  MIT Press.

Kim, J. (1999), `Making sense of emergence', {\em Philosophical Studies} {\bf 95}, pp. 3-36; 
reprinted in Bedau and Humphreys (2008); page reference to the reprint.

Kim, J. (2005), {\em Physicalism, or Something Near Enough},  Princeton University Press.

Kim, J. (2006), `Emergence: Core Ideas and Issues', {\em Synthese} {\bf 151}, pp. 547-559.

Klein, C. (2009), `Reduction without reductionism: a defence of Nagel on connectability', 
{\em Philosophical Quarterly} {\bf 59}, pp. 39-53.

Ladyman, J., and Ross, D. et al. (2007), {\em Every Thing Must Go: metaphysics naturalized}, 
Oxford University Press.

Lewis, D. (1970), `How to define theoretical terms',  {\em Journal of Philosophy} {\bf 67}, 
pp. 427-446; reprinted in his {\em Philosophical Papers Volume I}, Oxford University Press; 
page reference to the reprint.

Lewis, D. (1986) {\em On the Plurality of Worlds}, Oxford: Blackwell.

Loar, B. (1981), {\em Mind and Meaning},  Cambridge University Press.

McLaughlin, B. (1992), `The rise and fall of British emergentism'; in A. Beckerman, H. Flohr 
and J. Kim (eds.) {\em Emergence or Reduction?}, Berlin: de Gruyter; reprinted in Bedau and 
Humphreys (2008)

McLaughlin, B. (1997), `Emergence and supervenience', {\em Intellectica} {\bf 25} pp. 33-43, 
reprinted in Bedau and Humphreys (2008); page reference to the reprint.

Mainwood, P. (2006), {\em Is more different?}, Oxford D.Phil dissertation.

Marras, A. (2002), `Kim on reduction', {\em Erkenntnis} {\bf 57}, pp. 231-257.

Mostowski, A. (1966), {\em Thirty Years of Foundational Studies}, Oxford: Blackwell.

Nagel, E. (1961), {\em The Structure of Science: Problems in the Logic of Scientific 
Explanation}, Harcourt.

Nagel, E. (1979), `Issues in the logic of reductive explanations', in his {\em Teleology 
Revisited and other essays in the Philosophy and History of Science}, Columbia University 
Press; reprinted in Bedau and Humphreys (2008); page reference to the reprint.

Nagel, T. (1965), `Physicalism', {\em Philosophical Review} {\bf 74}, pp. 339-356.

Needham, P. (2009), `Reduction and emergence: a critique of Kim', {\em Philosophical Studies} 
{\bf 146}, pp. 93-116.

Needham, P. (2010), `Nagel's analysis of reduction: Comments in defence as well as critique', 
{\em Studies in History and Philosophy of Modern Physics} {\bf 41}, pp. 163-170.

Nickles, T. (1973), `Two concepts of inter-theoretic reduction', {\em Journal of Philosophy} 
{\bf 70}, pp. 181-201.

Papineau, D. (2010), `Can any sciences be special?', in {\em Emergence in Mind}
eds. C. Macdonald and G. Macdonald, Oxford University Press, pp. 179-197.

Paull, R. and Sider, T. (1992), `In defence of global supervenience', {\em Philosophy and 
Phenomenological Research} {\bf 52}, pp. 833-854.

Putnam, H. (1975), `Philosophy and our mental life', in his {\em Mind, Language and Reality}, 
Cambridge University Press, pp. 291-303.

Rueger, A. (2000), `Physical emergence, diachronic and synchronic', {\em Synthese} {\bf 124}, 
pp. 297-322.

Rueger, A. (2006), `Functional reduction and emergence in the physical sciences', {\em 
Synthese} {\bf 151}, pp. 335-346.

Scerri, E. (2007), `Reduction and emergence in chemistry---two recent approaches', {\em 
Philosophy of Science} {\bf 74}, pp. 920-931.

Schaffner, K. (1967), `Approaches to reduction', {\em Philosophy of Science} {\bf 34}, pp. 
137-147.

Schaffner, K. (1976), `Reductionism in biology: problems and prospects' in R. Cohen et al. 
(eds), {\em PSA 1974}, pp. 613-632.

Shapiro, L. (2000), `Multiple realizations', {\em Journal of Philosophy} {\bf 97}, pp. 
635-654.

Silberstein, M. (2001), `Converging on emergence: consciousness, causation and explanation', 
{\em Journal of Consciousness Studies} {\bf 8}, pp. 61-98.

Silberstein, M. (2002), `Reduction, emergence and explanation', in {\em The Blackwell Guide 
to the Philosophy of Science} eds. P. Machamer and M. Silberstein, pp. 80-107, Oxford: 
Blackwell.

Silberstein, M. and McGeever, J. (1999), `The Search for Ontological Emergence', {\em The 
Philosophical Quarterly} {\bf 49} pp. 182-200.

Simon, H. (1996), `Alternative views of complexity', Chapter 7 of his {\em The Sciences of 
the Artificial}, third edition, MIT Press; the Chapter is reprinted in Bedau and Humphreys 
(2008); page reference to the reprint.

Sklar, L. (1967), `Types of intertheoretic reduction', {\em British Journal for the 
Philosophy of Science} {\bf 18}, pp. 109-124.

Sober, E. (1999), `The multiple realizability argument  against reductionism',  {\em 
Philosophy of Science} {\bf 66}, pp. 542-564.

Stalnaker, R. (1996), `Varieties of supervenience', {\em Philosophical Perspectives} {\bf 
10}, pp. 221-241.

Teller, P. (1984), `A poor man's guide to supervenience and determination', {\em Southern 
Journal of Philosophy}, Supplement to volume {\bf 22}, pp. 137-162.

van Cleve, J. (1990), `Emergence vs panpsychism: magic or mind dust?' {\em Philosophical 
Perspectives} {\bf 4}, pp. 215-226.

\end{document}